\documentclass[pre,aps,superscriptaddress,twocolumn]{revtex4-1}
\usepackage{graphicx}
\usepackage[caption=false]{subfig}
\usepackage{epstopdf}
\usepackage{amsmath}
\usepackage[usenames,dvipsnames]{color}

\usepackage[normalem]{ulem}

\begin{document}

\title{Avalanches, thresholds, and diffusion in meso-scale amorphous plasticity}

\author{Botond Tyukodi}
\affiliation{Northeastern University, Department of Mechanical and Industrial Engineering \\ Boston, USA}
\author{Damien Vandembroucq}
\affiliation{PMMH, ESPCI Paris, CNRS UMR 7636, Sorbonne Universit\'e, Universit\'e Paris  Diderot, PSL Research University}
\author{Craig E Maloney}
\affiliation{Northeastern University, Department of Mechanical and Industrial Engineering \\ Boston, USA}

\pacs{--}

\begin{abstract}
We present results on a meso-scale model for amorphous matter in athermal, quasi-static (a-AQS), steady state shear flow.  In particular, we perform a careful analysis of the scaling with the lateral system size, $L$, of: i) statistics of individual relaxation events in terms of stress relaxation, $S$, and individual event mean-squared displacement, $M$, and the subsequent load increments, $\Delta \gamma$, required to initiate the next event; ii) static properties of the system encoded by $x=\sigma_y-\sigma$, the distance of local stress values from threshold; and iii) long-time correlations and the emergence of diffusive behavior.  
For the event statistics, we find that the distribution of $S$ is similar to, but distinct from, the distribution of $M$.  
The exponents governing the scaling properties of $P(S)$ completely determine the exponent $\alpha$ governing the finite size scaling of the load increment required to trigger the next event $\langle \Delta\gamma \rangle\sim L^{-\alpha}$.  
$P(M)$ is analogous to but distinct from $P(S)$.
We find a strong correlation between $S$ and $M$ for any particular event, with $S\sim M^{q}$ with $q\approx 0.65$.
This new exponent, $q$, completely determines the scaling exponents for $P(M)$ given those for $P(S)$.
For the distribution of local thresholds, we find $P(x)$ is analytic at $x=0$, and has a value $\left. P(x)\right|_{x=0}=p_0$ which scales with lateral system length as $p_0\sim L^{-a_1}$.  
In our model, by construction, the minimum, $x_{\text{min}}$, of $x$ in any particular configuration is precisely equal to $\Delta\gamma$, and, also by construction, $\langle S \rangle = \langle \Delta \gamma \rangle $.
Extreme value statistics arguments lead to a scaling relation between the exponents governing $P(x)$ and those governing $P(S)$.  
Finally, we study the long-time correlations via single-particle tracer statistics.
At short times, the displacement distributions are strongly non-Gaussian and consistent with exponentials as observed at short times in other driven and thermal glassy systems.  
At long times, a diffusive behavior emerges where the distributions become Gaussian.  
The value of the diffusion coefficient is completely determined by $\langle \Delta \gamma \rangle$ and the scaling properties of $P(M)$ (in particular from $\langle M \rangle$) rather than directly from $P(S)$ as one might have naively guessed.  
Our results: i) further define the a-AQS universality class with the identification of new scaling exponents unrelated to old ones, ii) help clarify the relation between avalanches of stress relaxation and long-time diffusive behavior, iii) help clarify the relation between local threshold distributions and event statistics and iv) should be important for any future work on the broad class of systems which fall into this universality class including amorphous alloys, glassy polymers, compressed granular matter, and soft glasses like foams, emulsions, and pastes. 
\end{abstract}

\date{\today}

\maketitle

\section{Introduction}
Amorphous solids such as amorphous alloys \cite{Antonaglia2014}, foams~\cite{Durian1997}, emulsions~\cite{Vasisht2018}, pastes~\cite{Basu2014,Seth2011}, colloidal glasses~\cite{Schall2007a}, granular materials~\cite{Papadopoulos2016, Hayman2011}, etc. exhibit avalanches of stress relaxation when driven slowly at low temperature.
This has been observed in experiments~\cite{Antonaglia2014, Denisov2016, Hayman2011} and computer simulations.
The computer simulations can be either of particulate~\cite{Lemaitre2009,Maloney2008} or meso-scopic nature~\cite{Talamali2011a,Lin2014a,Budrikis2017}.
In the meso-scopic models, space is broken into local regions, any one of which may suffer a yielding event, after which it must redistribute at least some portion of the stress it had been supporting.
The avalanches arise from cascades of local yielding events, so called shear transformations, which interact with each other elastically.

The distribution of avalanche sizes, $P(S)$, has been shown to exhibit critical scaling~\cite{Talamali2011a} as in other driven critical systems such as sand piles~\cite{Bak1987}, contact lines~\cite{Tyukodi2014} etc.
$P(S)$ is also related to other properties of the system beyond the spectrum of avalanches.
Recently, Lin and co-workers~\cite{Lin2014a} have argued that the avalanches and $P(S)$ place strong constraints on the form of the distribution, $P(x)$ of local residual stress, $x=\sigma_y-\sigma$, where $\sigma_y$ is the local threshold and $\sigma$ is the local stress.  
It is also well known that particulate computer simulations show an anomalous diffusion coefficient, $D$~\cite{Lemaitre2009,Maloney2008}, and Lemaitre and Caroli have argued~\cite{Lemaitre2009} that the size dependence of $D$ can be understood in terms of the geometrical properties of the deformation; in particular how avalanches organization into lines in two dimensions (2D).
We have recently shown~\cite{Tyukodi2018} that the meso-scopic models also show an anomalous system size dependent diffusion coefficient similar to that observed in particulate simulations.
However, in either case, the precise connection between the scaling properties of diffusive quantities and the avalanche spectrum has not been studied.
In the present paper, we perform a careful finite-size scaling analysis simultaneously on: i) the distribution of event sizes, ii) the distribution of local residual stress and iii) the single particle displacement statistics at longer time.

For the event size distribution, we characterize individual events both in terms of the stress released, $S$, and also in terms of the mean squared displacement (MSD), $M$.
We find a surprising, non-trivial relation between the individual event stress relaxation, $S$, and its MSD, $M$ with $M\sim S^{q}$ with $q\approx 0.65$.
We do not provide any deep understanding on the origin of this relation or the value of the new scaling exponent, however, we show that it completely determines the form of the distribution $P(M)$ from the distribution $P(S)$.
We also study the distribution of load increments, $\Delta\gamma$, effectively like the waiting time between events of any size, and show that it is essentially exponential with an average $\langle \Delta \gamma \rangle$ equal to $\langle S \rangle$ as it must be in steady state.  

For the thresholds embodied in the $P(x)$ distribution, we agree with Lin {\it et. al.} that the scaling of $\langle x_{min} \rangle=\langle \Delta \gamma \rangle=\langle S \rangle$ is consistent with what one would obtain from the minima of uncorrelated samples of $P(x)$.
However, the form of our $P(x)$ distribution is qualitatively different than what was found by Lin {\it et. al.}.
Lin {\it et. al.} found a power-law form for $P(x)$ and point out the distinction with other depinning systems where $P(x)$ is analytic. 
Here, we will show that $P(x)$ is actually analytic at $x=0$, but with the value of $P(x)$ at $x=0$ scaling in a non-trivial way with $L$ such that the extreme value statistics prediction for $\langle x_{min} \rangle$ based on $P(x)$ is consistent with our explicit measurement of it.

Finally, for the diffusive behavior, in agreement with our earlier results~\cite{Tyukodi2018}, we find an anomalous size dependence, $D~\sim L^{1.05}$.
One might have naively expected $D\sim L^{d-d_f}$ where $d_f$ is the fractal dimension one would infer from the $P(S)$ distribution.
Our data is inconsistent with this naive expectation.
However, we find that taking into account the non-trivial relation between $S$ and $M$ for individual events, one simply finds that $D=\langle M \rangle / \langle \Delta \gamma \rangle$.
Thus the $P(M)$ distribution along with the $q$ exponent completely determine the size dependence of $D$.

\section{A mesoscopic model of amorphous plasticity}\label{section:Model}
We use a coarse-grained, depinning-like lattice model of amorphous plasticity (for a recent review of such mesomodels see \cite{Nicolas2017, Rodney2011}). These models provide a semi-continuous description, preserving the two key ingredients of amorphous plasticity: the elastic interations between the shear transformations and the disordered potential landscape. Shear transformations are ``replaced'' by Eshelby inclusions and the disorder is introduced via activation stress barriers. In what follows, we provide some insight into mesomodels.
\subsection{Elastic interactions: Eshelby}
\begin{figure}[ht]
\begin{center}
\subfloat[]{
\includegraphics[width=3.5cm]
{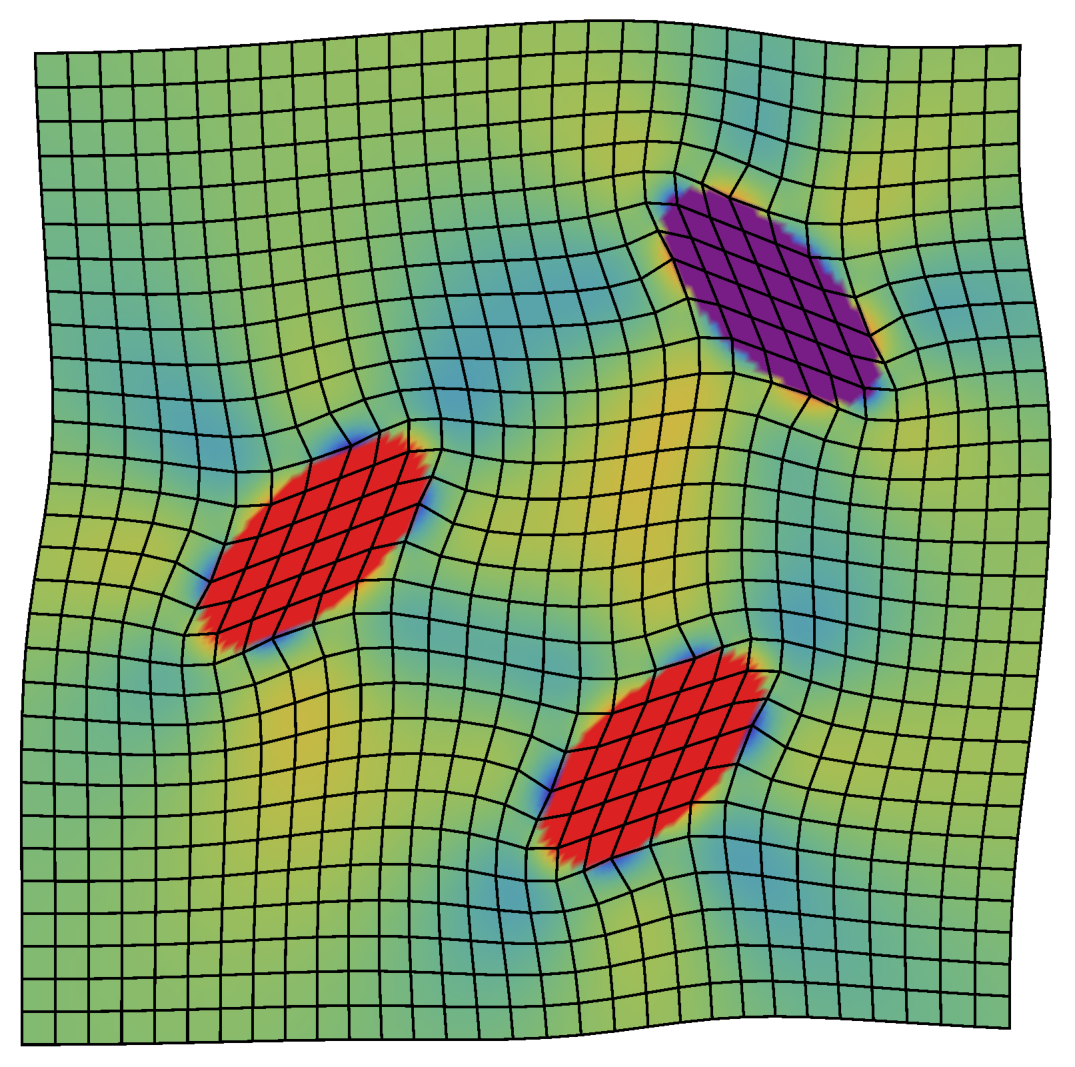}
}
\subfloat[]{
\includegraphics[width=4.5cm]
{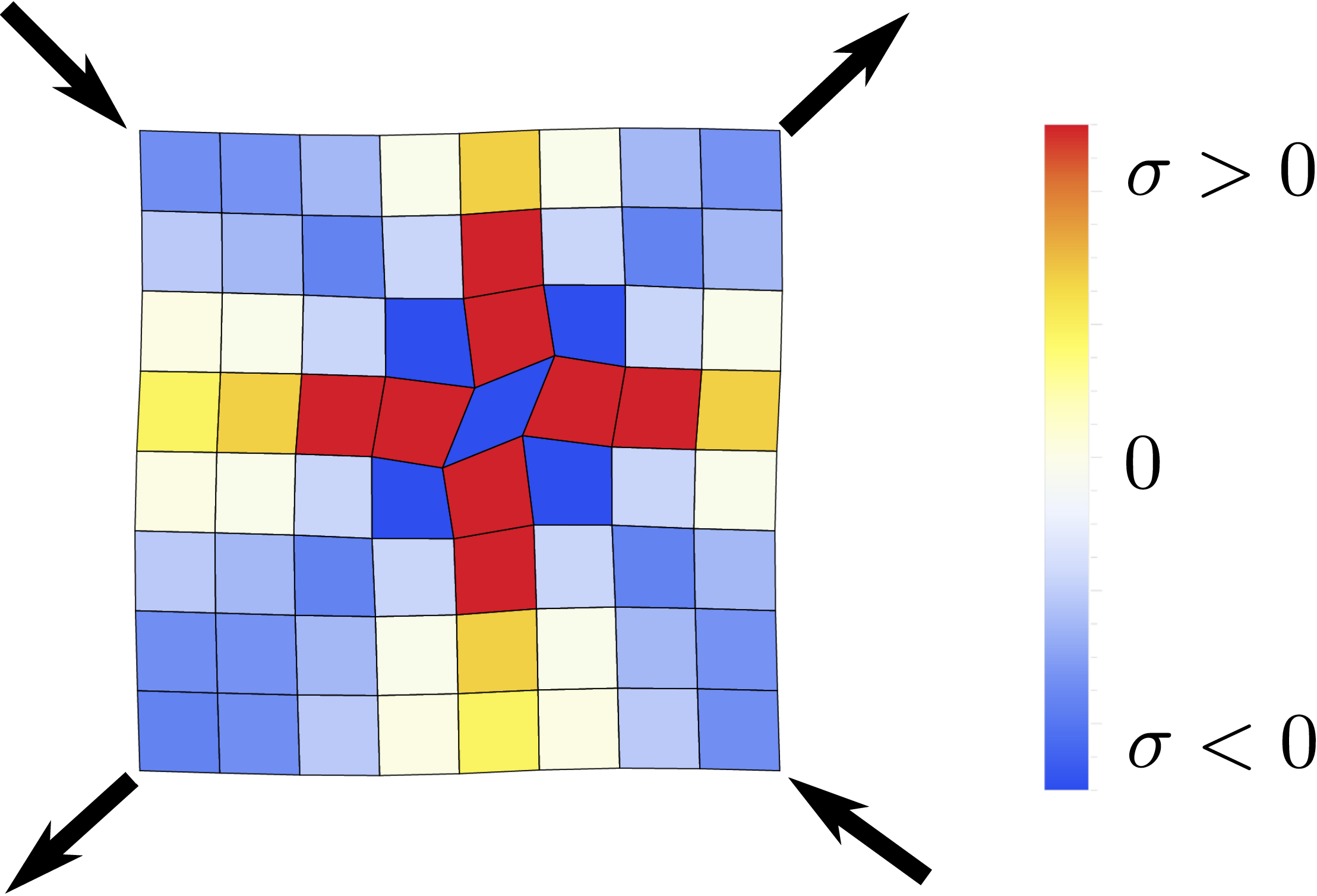}
}
\caption{\label{fig:eshelby} (a) Interacting Eshelby inclusions in an elastic bulk. (b): discretized kernel associated to simple shear deformations.}
\end{center}
\end{figure}

Mesomodels attempt to preserve the elastic interaction between shear transformations upon coarse graining. Shear transformations are therefore replaced by their continuous counterpart of material inclusions known as Eshelby inclusions. These inclusions have the same elastic properties as the material, however, they can undergo permanent deformation, just as the shear transformations do. Fig. \ref{fig:eshelby} shows such interacting inclusions.


Here we consider a scalar, two dimensional description where the plastic deformation of the inclusions obeys the symmetry of the external loading. In other words, a pure shear loading $\sigma = \sigma_{xx}-\sigma_{yy}$ can only cause a plastic strain $\epsilon^p = \epsilon^p_{xx} - \epsilon^p_{yy}$, while a simple shear loading of $\sigma=\sigma_{xy}$ can only result in a plastic strain of the form $\epsilon^p = \epsilon^p_{xy}$. From here on we work in the simple shear geometry.

When an inclusion undergoes a plastic deformation, it induces an elastic field in the rest of the bulk. The elastic field is the solution of the Eshelby inclusion problem \cite{Eshelby1957a} and it is known that its far-field solution is independent of the inclusion's shape. In two dimensions, the far-field solution of the Eshelby inclusion problem resulting from a plastic deformation $\epsilon^p$ gives an induced stress of the form $G(r, \theta) \sim \epsilon^p \cos(4 \theta)/r^2$. The stress induced by an inclusion thus has a quadrupolar symmetry as shown on Fig. \ref{fig:eshelby} and has positive signs along certain directions and negative signs along others. The triggering of further rearrangements is therefore favored along the positive directions, but the material is stabilized along the negative ones.

Note that the elastic kernel $G(r, \theta) \sim 1/r^2$ in two dimensions, hence it is long ranged. Therefore, discretizing the elastic fields induced by the Eshelby inclusions in periodic geometries is not a trivial task. There are various discretization schemes to discretize the Eshelby fields on a square lattice, in periodic gemoetries such as Fourier-space discretization \cite{Talamali2011, Tyukodi2016a} or finite element methods \cite{Nicolas2014, Tyukodi2016a}. Here we chose a FE scheme as shown on Fig. \ref{fig:eshelby}. Details regarding the discretization are presented in the Appendix of ref. \cite{Tyukodi2018}, here we just mention that the finite element kernels give realistic near field interactions. The Eshelby inclusions associated to shear transformations then live on square lattice sites. Once a site deforms plastically, the stress is redistributed over the system according to the discretized version of $G$.

\subsection{Disorder: distributed thresholds vs distributed slip amplitudes}
\
\begin{figure}[ht]
\begin{center}
\includegraphics[width=6cm]
{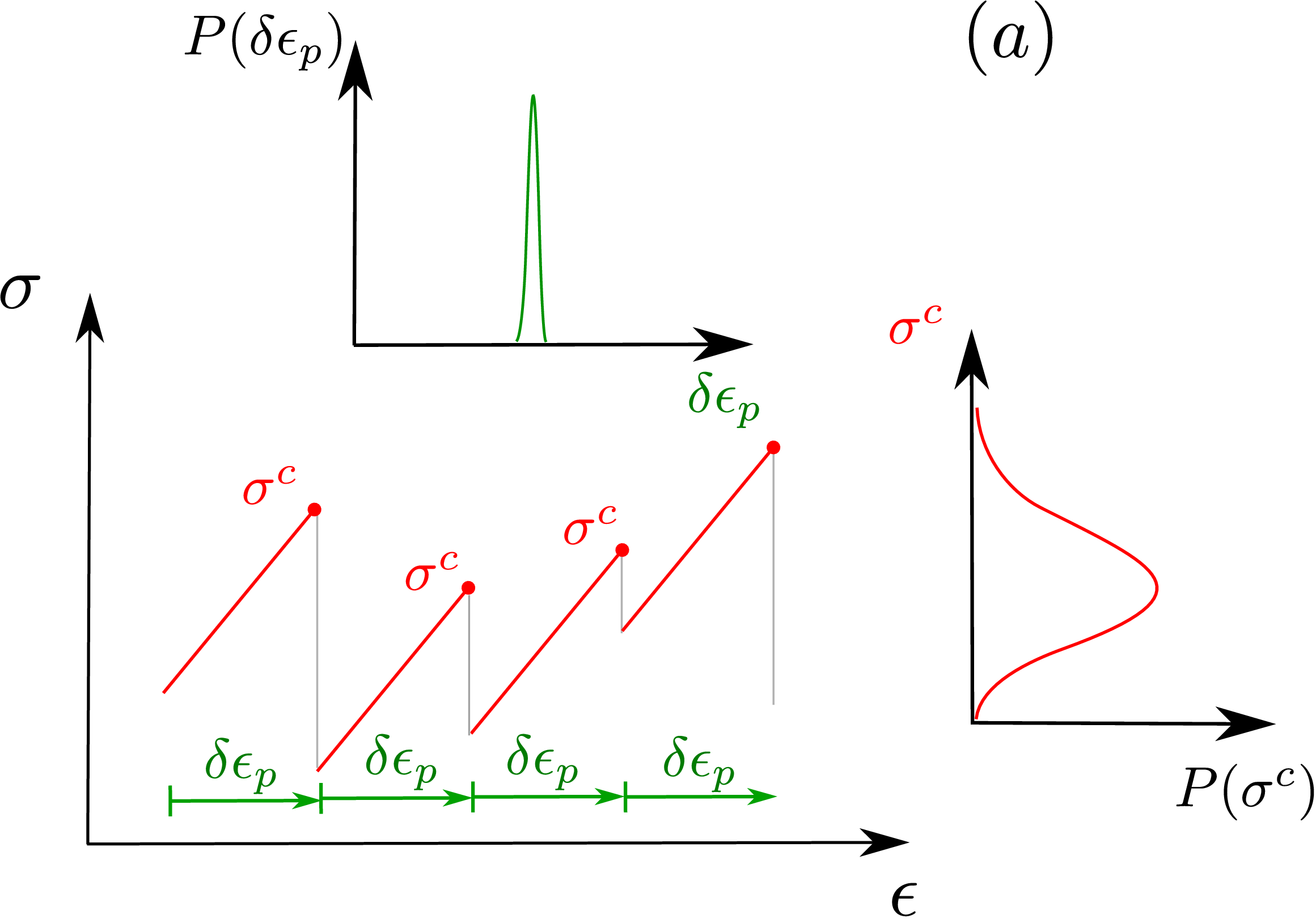}
\includegraphics[width=6cm]
{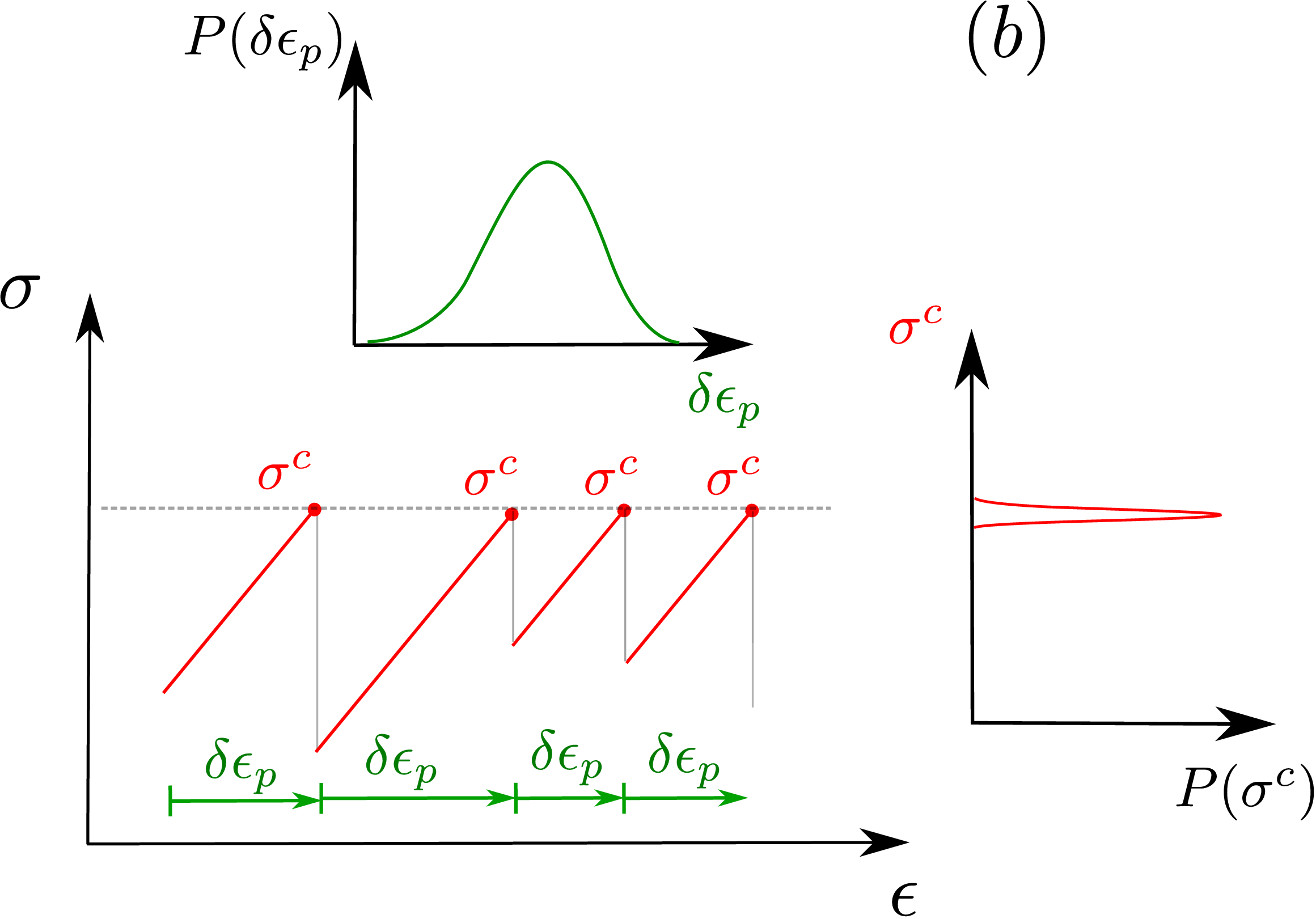}
\caption{\label{fig:disorder} Disorder enters into the model either (a) via random activation thresholds or (b) random slip increments. The main plots show the stress landscape experienced by one of the discrete elements, whereas the side plots show the corresponding threshold and slip increment distributions.}
\end{center}
\end{figure}

The activation of shear transformations is related to energy or stress barriers. Here we consider an athermal model, therefore the plasticity of inclusions is governed by local stress barriers: whenever the stress on a particular site exceeds a threshold value $\sigma^c$, the site slips some amount $\epsilon^p$. The structural disorder of the material is then reflected in the distributions and correlations of $\sigma^c$ and $\epsilon^p$. Here we consider two particular forms of the disorder: either $\sigma^c$ is uniformly distributed from $[0.5,1.5)$ and $\epsilon^p = \epsilon_0 / 2$ or $\sigma^c = 1$ and $\epsilon^p$ uniformly distributed from $[0, \epsilon_0)$ with $\epsilon_0=1$. We name the distributed threshold model Y1 and the distributed slip increment model Y0. Recalling the depinning analogy, Y1 introduces disorder as fluctuations of the depths of the potential wells, while Y0 accounts for disorder as the fluctuations of the widths of the wells as shown in Fig. \ref{fig:disorder}. We were experimenting with the combination of the 
two protocols (random initial stresses with constant thresholds and constant slip increments or random initial plastic strains with the associated residual stresses with constant thresholds and constant increments), these however resulted in a single narrow and persistent 
shear band, consistent with \cite{Rodney2011}, showing that disorder in the initial conditions only is not enough.

We choose a quasistatic, strain driven driving as follows: the strain of the system is increased up until the point one of the sites yield. The strain is then held constant as long as further events are triggered and only after all events stopped is adjusted again to trigger the next one. This loading allows for true quasistatic loading without a finite strain step.

\section{Elementary events: avalanches}\label{section:Avalanches}
\begin{figure}[ht]
\begin{center}
\includegraphics[width=6cm]
{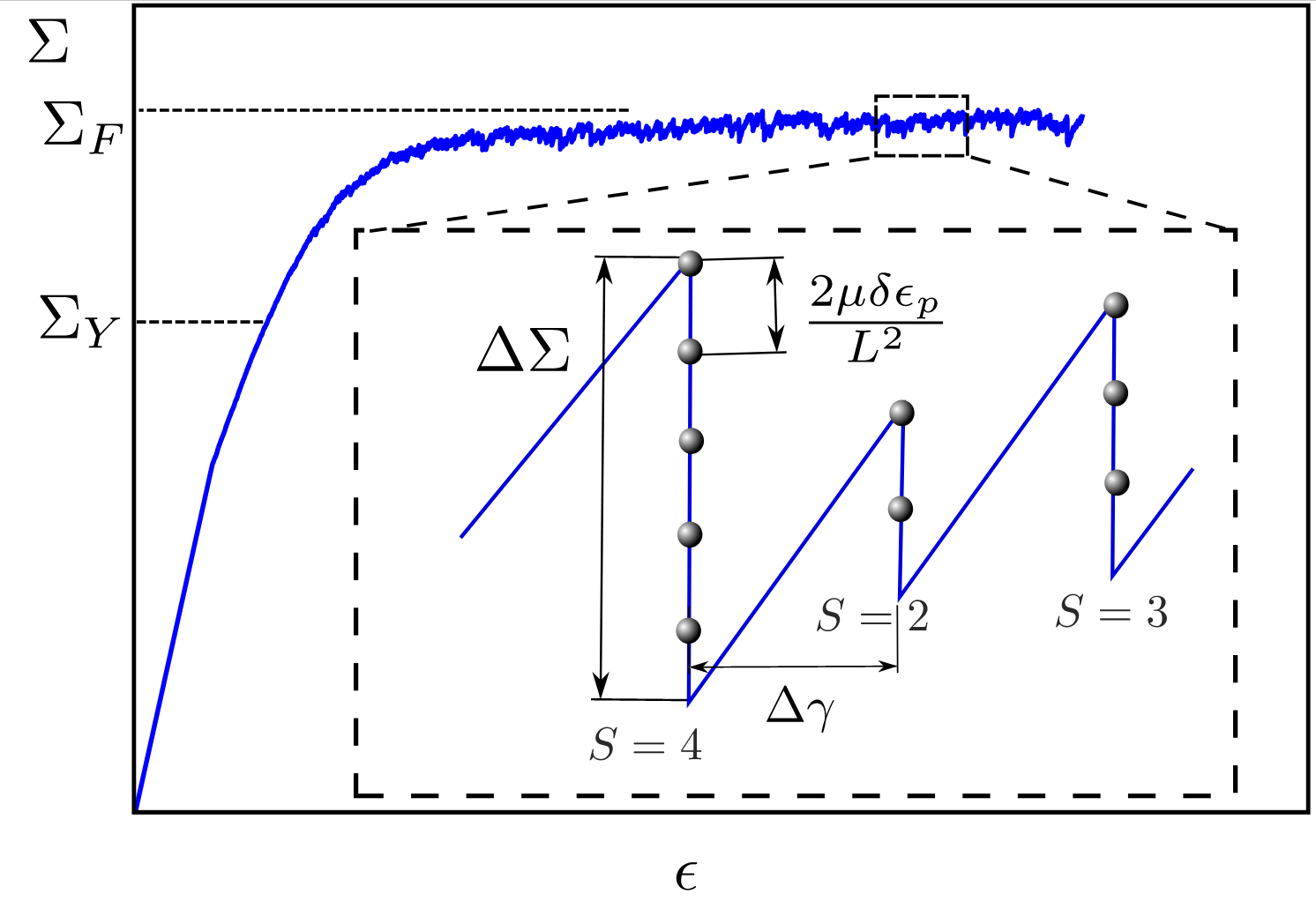}
\caption{\label{fig:stress_strain} Serrated flow curve. Fluctuations of the stress (i.e. the sudden stress drops) correspond to individual avalanches. The stress (strain) increase between successive avalanches is $2 \mu \Delta \gamma$ ($\Delta \gamma$). }
\end{center}
\end{figure}

The flow of amorphous materials is characterized by intermittent dynamics resulting in sudden stress drops in the flow curve (Fig. \ref{fig:stress_strain}). These drops correspond to the collective activation of many shear transformations so plasticity happens in terms of bursts known as avalanches. 
In this section first we investigate the size distribution of such elementary events in terms of the stress drop.
We then study the distribution of subsequent load increments and show that it is exponential with a characteristic scale determined completely by the average stress drop. 
Finally, we study the individual event MSDs, showing that there is a strong correlation for any given event between its stress drop and its MSD.
This allows us to map the distribution of stress drops, $P(S)$, to the distribution of MSDs, $P(M)$.
We will show below that \emph{both} of these distributions are required to determine the diffusion coefficient.

\subsection{Stress drops}
Cascades are usually power-law distributed around the yielding transition and their upper cutoff's finite size scaling gives valuable information about the spatial structure of the avalanches. The normalized distribution of avalanches however looses an important information when it comes to size dependence, namely the cummulative number of avalanches in a given strain window. We therefore define the avalanche rate $R(S, L)$ (rather than the avalanche distribution) as the number of events of size $S$ per unit loading strain \cite{Salerno2012a} as $R(S,L) = n(S, S+\mathrm{d}S, \Delta \epsilon) / \mathrm{d}S \Delta \epsilon$, where $n(S, S+\mathrm{d}S, \Delta \epsilon)$ is the number of avalanches of size between $[S, S+\mathrm{d}S)$ within a strain window $\Delta \epsilon$. $L$ is the linear size of the system, so if the lattice constant is $a$, there are $(L/a)^2$ sites in the system. The avalanche size $S$ is defined as $S =  L^2  \Delta \Sigma/ 2 \mu$ as shown in Fig. \ref{fig:stress_strain}. This definition is equivalent to that of particle simulations \cite{Salerno2012a} and in the mesomodel, in the case of uniform slip increments $S$ would simply give the number of plastic events (flips) within the avalanche. Normalizing $R(S, L)$ then gives the usual probability distribution $P(S, L)$ of the avalanche sizes \cite{Lin2014a}.
In particle simulations \cite{Salerno2012a} it was found that $R(S, L)$ obeys the scaling
\begin{equation}\label{eq:R_SL}
 R(S, L) = L^\beta g(S/L^{d_f})
\end{equation}
with $\beta = 0.2 \pm 0.1$ for an overdamped system. $d_f$ is known as the fractal dimension of the avalanche and is a characteristic exponent of its spatial structure. The scaling function $g(y)$ is such that it recovers the power law $g(y) \sim y^{-\tau}$ for $y \ll 1$. For $S \ll L$ therefore we have $R(S, L) \sim L^\beta S^{-\tau}/L^{-d_f \tau}$. Well below the cutoff we then have $R(S, L) \sim L^\gamma S^{-\tau}$ with the scaling relation $\gamma = \beta + d_f \tau$.

Another scaling relation can be obtained by computing the cummulative event number within a unit strain window:
\begin{eqnarray}
S_{cumm} &=& \int_0^{\infty} S R(S,L) dS \nonumber \\
&=& L^{2 d_f + \beta} \int_0^{\infty} u g(u) \mathrm{d}u \sim L^{2 d_f + \beta}
\end{eqnarray}
On the other hand, in the steady state, the stress cannot increase nor decrease on average thus no elastic strain can be accummulated. This means that in steady state, on average, all the energy is dissipated and all the accumulated strain is plastic strain. Then in a strain window $\Delta \epsilon$ the cummulative number of events is given by $S_{cumm} =  L^2 \Delta \epsilon/\epsilon_0$ where $\epsilon_0$ is the typical plastic strain within an inclusion. Projected to a unit strain window $\Delta \epsilon=1$, $S_{cumm} \sim L^2$ and we arrive to the scaling relation $2 d_f + \beta = 2$.

\begin{figure}[ht]
\begin{center}
\includegraphics[width=8cm]
{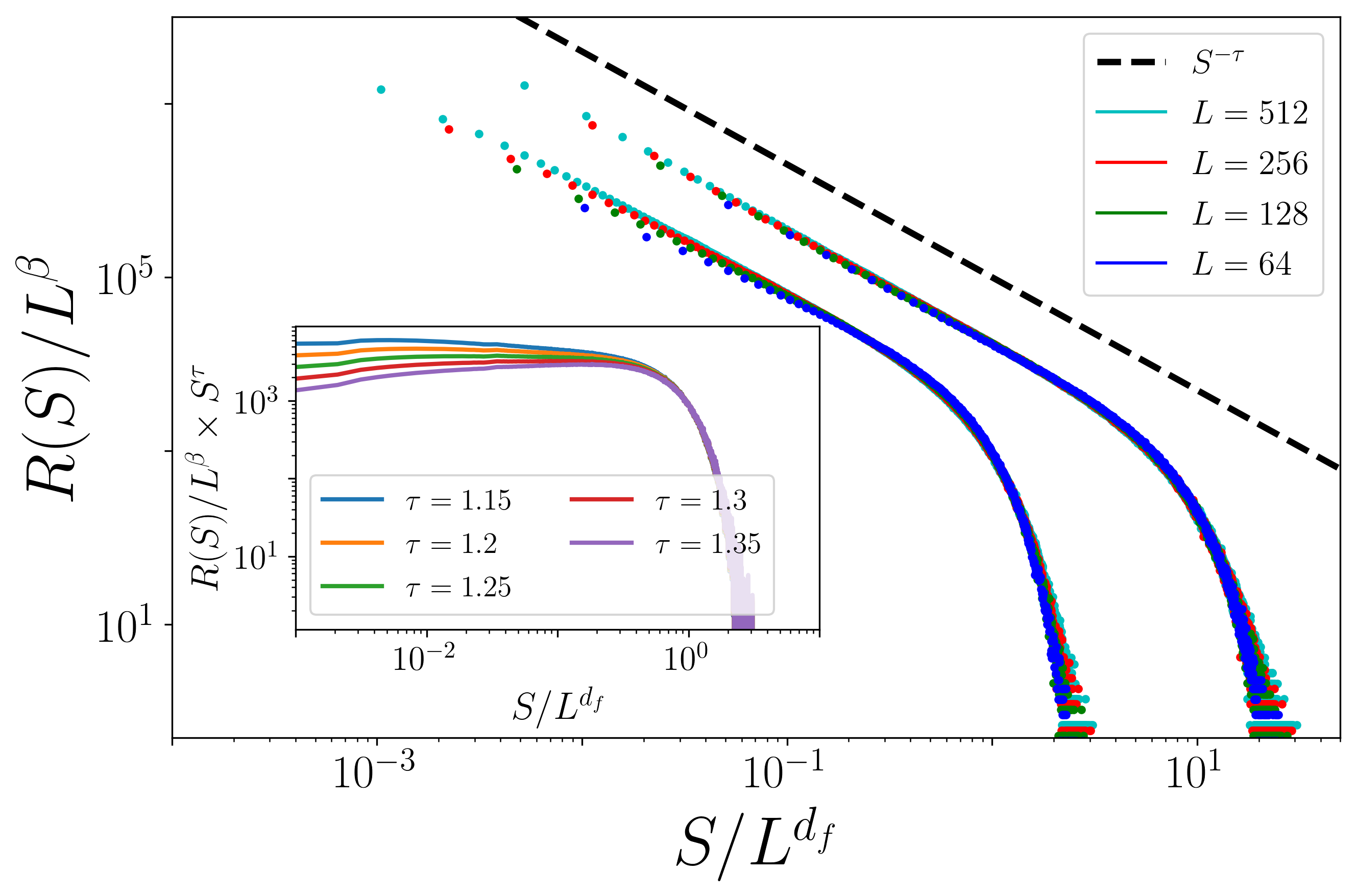}
\caption{\label{fig:avalanches} Avalanche rates rescaled with the system size, for the two disorder types. For clarity, plots of Y0 and Y1 were shifted apart horizontally. The left plot corresponds Y0, the right one to Y1. The dashed line is a guide for the eye with $\tau=1.3$. All models are well described by the same set of exponents: $\tau \approx 1.3, \beta \approx -0.2, d_f \approx 1.1$. Inset: Avalanche rates of the Y1 model flattened by various exponents $\tau$. Figure is intended to give an idea about the accuracy in the measurement of $\tau$.}
\end{center}
\end{figure}

Figure \ref{fig:avalanches} shows the rescaled avalanche rate $R(S, L) / L^\beta$ as a function of the rescaled avalanche size $S/L^{d_f}$ for various system sizes and the two kernels. We find an excellent collapse for the various system sizes, independently of the kernel or the type of disorder. Moreover, the values of the exponents $\tau, \beta, \gamma, d_f$ are also robust for both disorder types. For the avalanche exponent we find $\tau \approx 1.3 \pm 0.05$ which is very close to the value obtained from previous particle simulations \cite{Salerno2012a} $\tau=1.25$, Durian model simulations \cite{Maloney2015} $\tau=1.2$ and other mesomodels $\tau=1.3 \pm 0.05$ \cite{Lin2014, Budrikis2013a, Budrikis2017, Liu2016}, however considerably less than the mean field value $\tau=1.5$ \cite{Dahmen2011, Sethna2001, Dahmen2009}. For the fractal dimension $d_f$ we find a value $d_f =1.1$ which is larger than the value $d_f=0.9$ reported in particle simulations \cite{Salerno2012a} or $d_f=1$ obtained in a similar lattice model with extremal dynamics \cite{Talamali2011a}, however the same value was reported in other lattice models \cite{Lin2014}. For both disorder types we find $\gamma=1.25 \pm 0.05$ which is again close to the molecular dynamics value $\gamma=1.3 \pm 0.05$ \cite{Salerno2012a}. For the exponent $\beta$ we find $\beta=-0.2$ which is considerably different from the particle result $\beta=0.2$. Note that previous lattice models \cite{Lin2014, Talamali2011, Talamali2011a} focused on the normalized avalanche distribution $P(S)$, hence did not have access to the $\beta$ and $\gamma$ exponents. Note furthermore that the two scaling relations involving $\tau, \beta, \gamma, d_f$ are verified by our measured values.

A finite size scaling was revealed by Lin et al. \cite{Lin2014} by considering the normalized avalanche distribution $P(S, L)$ having the form 
\begin{equation}\label{wyart_avalanche}
P(S, L) \sim S^{-\tau} f(S/S_c)
\end{equation}
where $S_c$ is the upper cutoff of the avalanches and scales with the system size as $S_c \sim L^{d_f}$. The average avalanche size $\langle S \rangle$ can be computed and one finds $\langle S \rangle \sim L^{d_f(2-\tau)}$. The average stress drop $\langle \Delta \Sigma_{drop} \rangle$ is then given by $\langle \Delta \Sigma_{drop} \rangle = \langle S \rangle / L^d \sim L^{-\alpha}$ with $\alpha=d_f(2-\tau)-d$. Inserting $d_f=1.1$, $\tau=1.3$ and $d=2$ one obtains $\alpha=1.23$. 

\begin{figure}[ht]
\begin{center}
\includegraphics[width=8cm]
{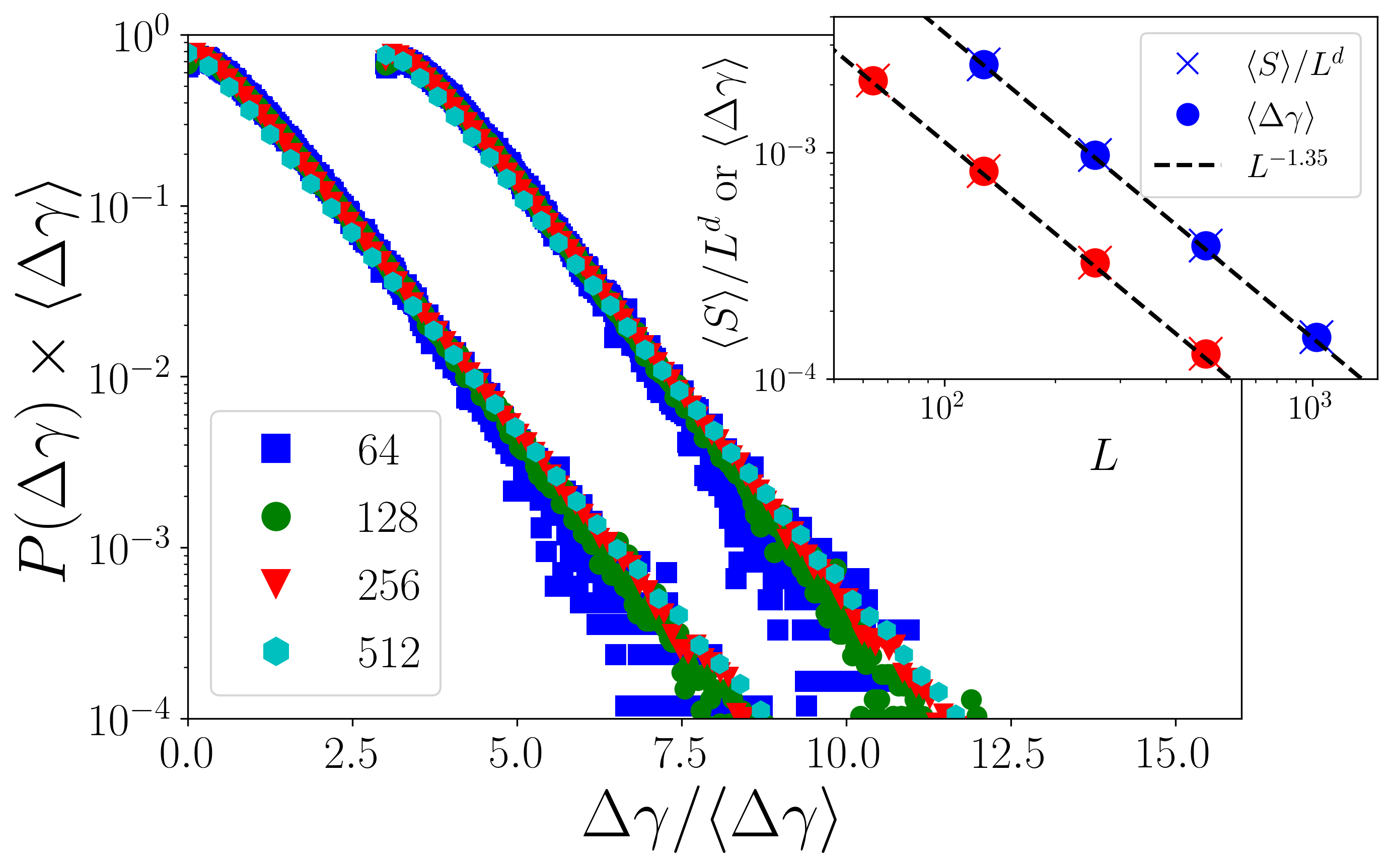}
\caption{\label{fig:avg_increase} Load increment distribution between avalanches for Y0 and Y1. $P(\Delta \gamma)$ distributions follow a simple waiting time distribution with a mean $\langle \Delta \gamma \rangle \sim L^{-\alpha}$ for both models. For readability, curves for different models were shifted horizontally and are Y0 (left), Y1 (right). Inset: average stress drop $\langle S \rangle / L^d$ (crosses) and average strain increment $\langle \Delta \gamma \rangle$ (filled circles) as a function of the system size, for the two models. Note that these two must be equal in steady state by construction in this family of models. To avoid overlap, data for Y0 and Y1 were shifted and are: Y0 (left), Y1 (right). The dashed line shows our best fit, $\langle S \rangle / L^d \sim L^{-\alpha}$ with $\alpha=1.35$. The infinite size scaling prediction $ \langle S \rangle / L^d \sim L^{d_f(2-\tau)-d}$ provides a smaller value, $d_f(2-\tau)-d = -1.23$.} 
\end{center}
\end{figure}

As shown on Fig. \ref{fig:avg_increase} inset, this prediction is close to the simulation, but does not precisely match it: our measured $\alpha$ is slightly larger, $\alpha=1.35$. This latter value matches results by Lin et. al.\cite{Lin2014}. The discrepancy between the predicted value $\alpha=1.23$ and the measured one $\alpha=1.35$ may stem from the fact that the distribution function in eq. \ref{wyart_avalanche} is not properly normalized. A proper normalization would involve knowledge about the lower cutoff $S_m$ of the power law $P(S,L)$. Eq. \ref{wyart_avalanche} does not include such a normalization factor, therefore corrections may arise.

\begin{figure}[ht]
\begin{center}
\includegraphics[width=8cm]
{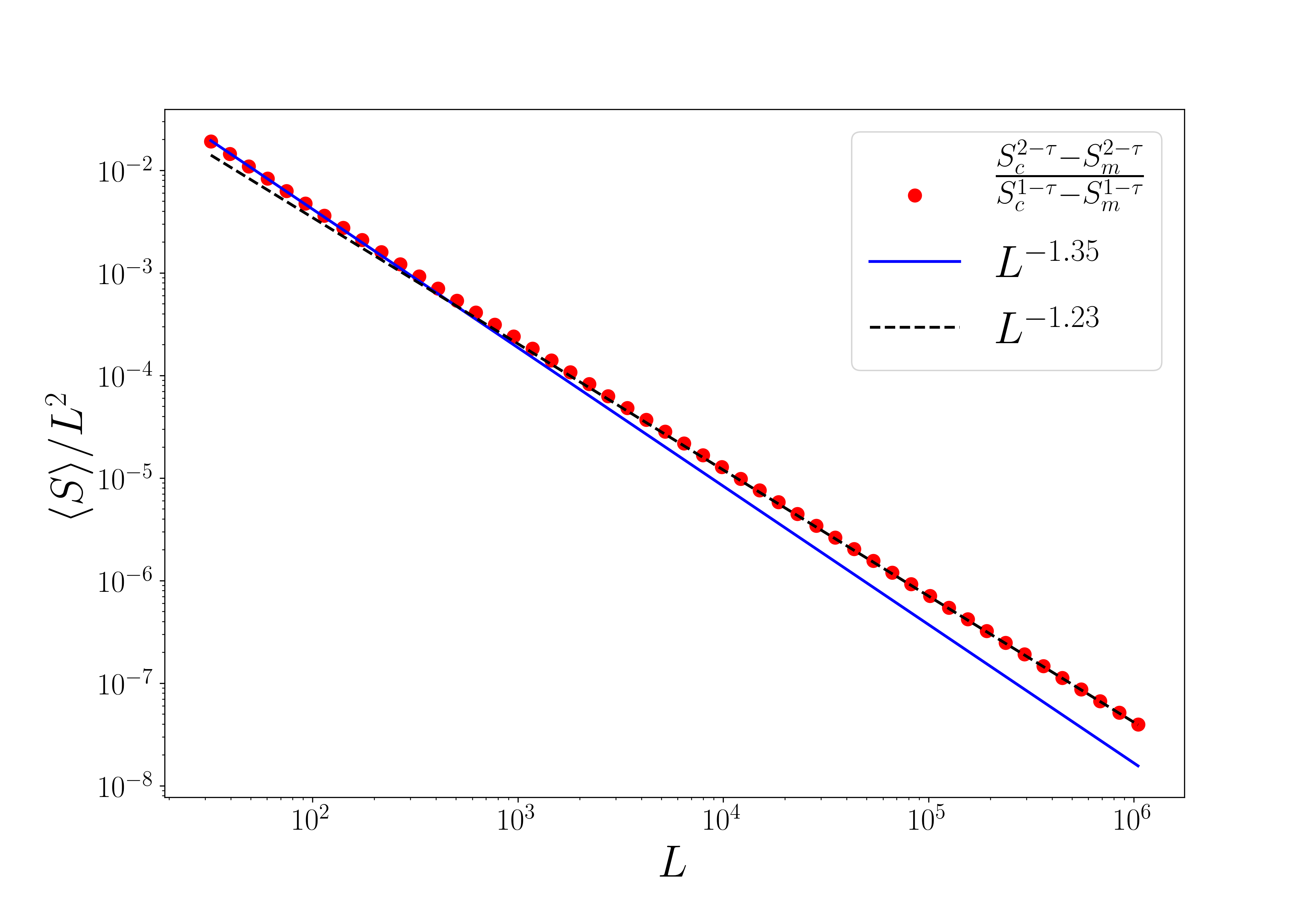}
\caption{\label{fig:s_mean_scale} Apparent scaling at intermediate $L$. Red dots indicate the mean avalanche size estimate from eq.  \ref{eq:mean_av}, considering a lower cutoff value $S_m=1.0$. The infinite-size scaling is visible only above very large systems, $L>2000$.}
\end{center}
\end{figure}

To show the effect of the lower cutoff, we compute $\langle S \rangle$ from the unnormalized avalanche rate $R(S, L)$, approximating the bounds by a hard lower cutoff $S_m$ and a hard upper cutoff $S_c \sim L^{d_f}$:
\begin{equation}\label{eq:mean_av}
 \langle S \rangle = \frac{\int_{S_m}^{S_c} S R(S, L) dS}{ \int_{S_m}^{S_c} R(S, L) dS} \sim \frac{S_c^{2-\tau} - S_m^{2-\tau}}{S_c^{1-\tau} - S_m^{1-\tau}}
\end{equation}
If $1 < \tau < 2$, at $L \to \infty$, the leading term in the numerator is $S_c^{2-\tau}$ and in the denominator, $S_m^{1-\tau}$. We find:
\begin{equation}
  \langle S \rangle \sim \frac{S_c^{2-\tau}}{S_m^{1-\tau}}
\end{equation}
Recall that $S_c \sim L^{d_f}$. Assuming that $S_m$ is size-independent, we recover $\langle S \rangle \sim L^{d_f(2-\tau)}$ at the thermodynamic limit. At intermediate system sizes, however, $S_m$ cannot be neglected. Figure \ref{fig:s_mean_scale} shows that the presence of a lower cutoff gives rise to an apparent scaling at intermediate system sizes with $\alpha=1.35$.

\subsection{Load increments}


In figure~\ref{fig:avg_increase}, we plot the distribution of loading increments, $\Delta\gamma$, required to trigger new events.
We can think of this equivalently as the inter-event waiting time distribution.
Since, in steady state, all loading increments must be offset by load drops occurring in avalanches, we must have that the average intensive stress drop is equal to the average load increment: $\langle S \rangle / L^d = \langle \Delta \gamma \rangle$.
Figure~\ref{fig:avg_increase} shows that the load increments follow a waiting time distribution and they are essentially distributed exponentially. The inset shows that indeed $\langle S \rangle / L^d = \langle \Delta \gamma \rangle$ and verifies the finite size scaling of $\langle S \rangle / L^d$ or $\langle \Delta \gamma \rangle$ as predicted by eq. \ref{eq:mean_av}. Insets of Fig. \ref{fig:individual_avalanches} show that, for individual events, $S$ and $\Delta \gamma$ are uncorrelated, further supporting that $\Delta \gamma$ follows a waiting time statistics. Even though avalanches consist of highly correlated events, the inter-event load increments resemble simple waiting times.


\subsection{Individual event mean-squared displacement}
\begin{figure}[h] 
\includegraphics[width=8cm]{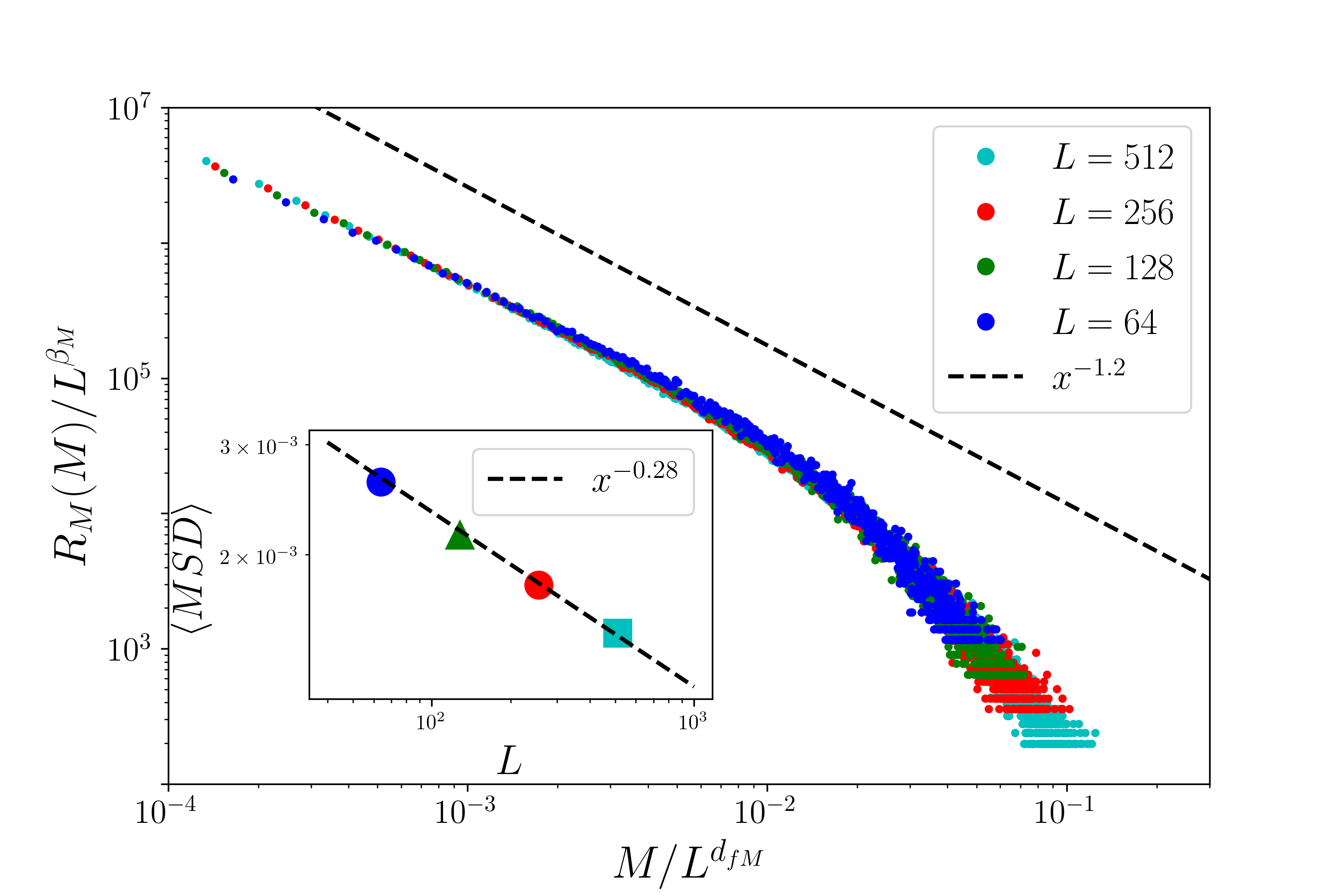}\\
\includegraphics[width=8cm]{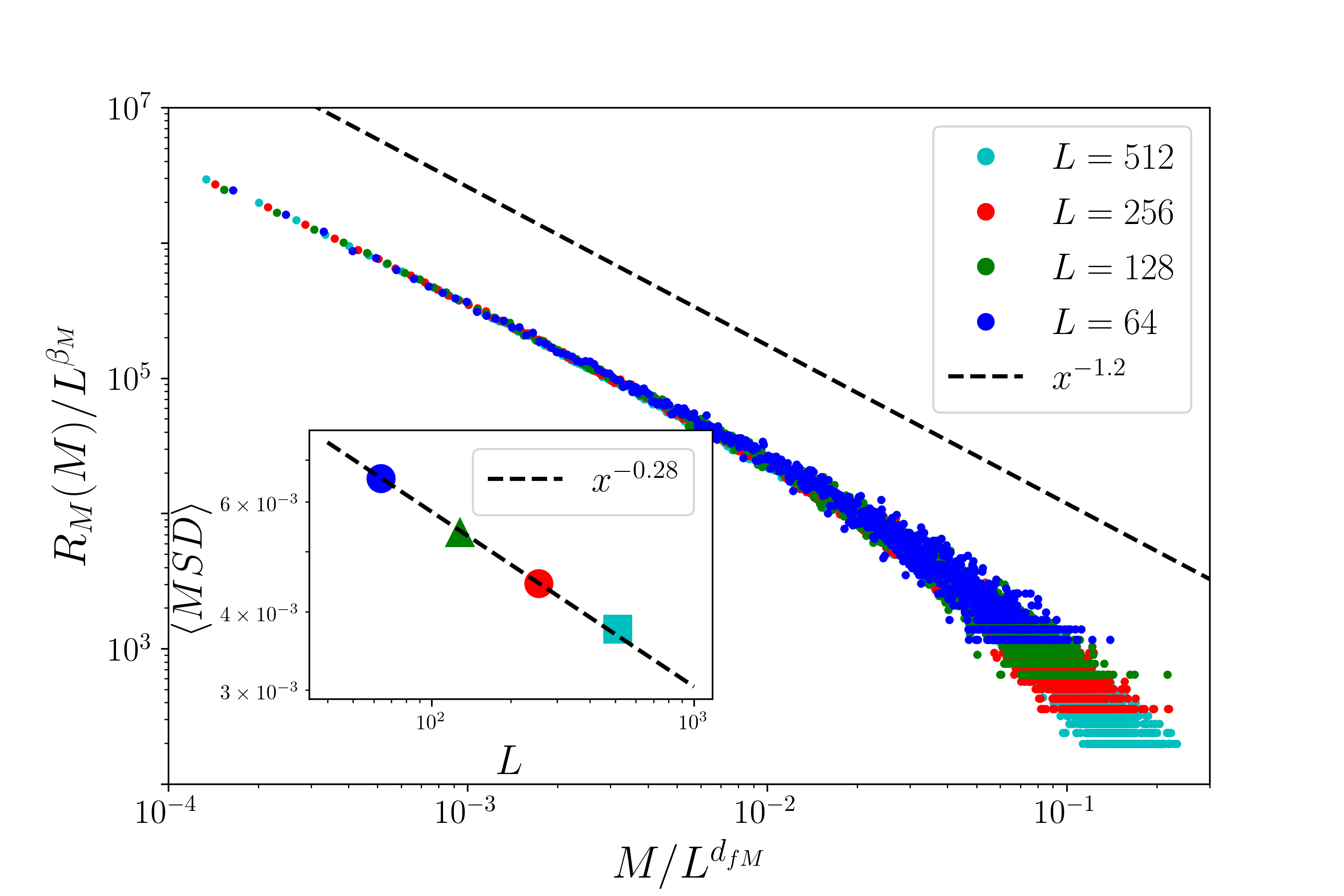}
\caption{\label{fig:P_MSD} Distribution of mean square displacements in individual avalanches. Inset: the average per-avalanche mean square displacement as a function of system size: $\langle MSD \rangle \sim L^{-0.28}$. Top: Y0, bottom: Y1. }
\end{figure}  
Similarly to avalanche sizes, one can measure the accumulated mean square displacement during an avalanche. We find a similar scaling ansatz for the rates of mean square displacements, namely:
\begin{equation}
 R_M(M, L) = L^{\beta_M} g_M(M/L^{d_{fM}})
\end{equation}
Figure \ref{fig:P_MSD} shows data rescaled according to the above ansatz. The function $g_M(x)$ is such that $g_M(x) \sim x^{-\tau_M}$ for $x \ll 1$. Values of the $\beta_M, d_{fM}, \tau_M$ exponents along with their avalanche size counterparts are summarized in table \ref{tab:exponents}.

\begin{table}[]
\begin{tabular}{|l|l|l|l|l|l|l|l|l|}
\hline
     & $\tau$ & $d_f$ & $\beta$ & q    & $\tau_M$ & $d_{fM}$ & $\beta_M$ & $\gamma$ \\ \hline
Y0 & 1.3    & 1.1   & -0.2    & 0.65 & 1.2      & 0.1      & 0.85  & 1.25    \\ \hline
Y1 & 1.3    & 1.1   & -0.2    & 0.65 & 1.2      & 0.1      & 0.85  & 1.25     \\ \hline
\end{tabular}\caption{\label{tab:exponents} Exponents measured. Precision on all exponents is $\pm 0.05$}
\end{table}

\begin{figure}[h] 
\includegraphics[width=8cm]{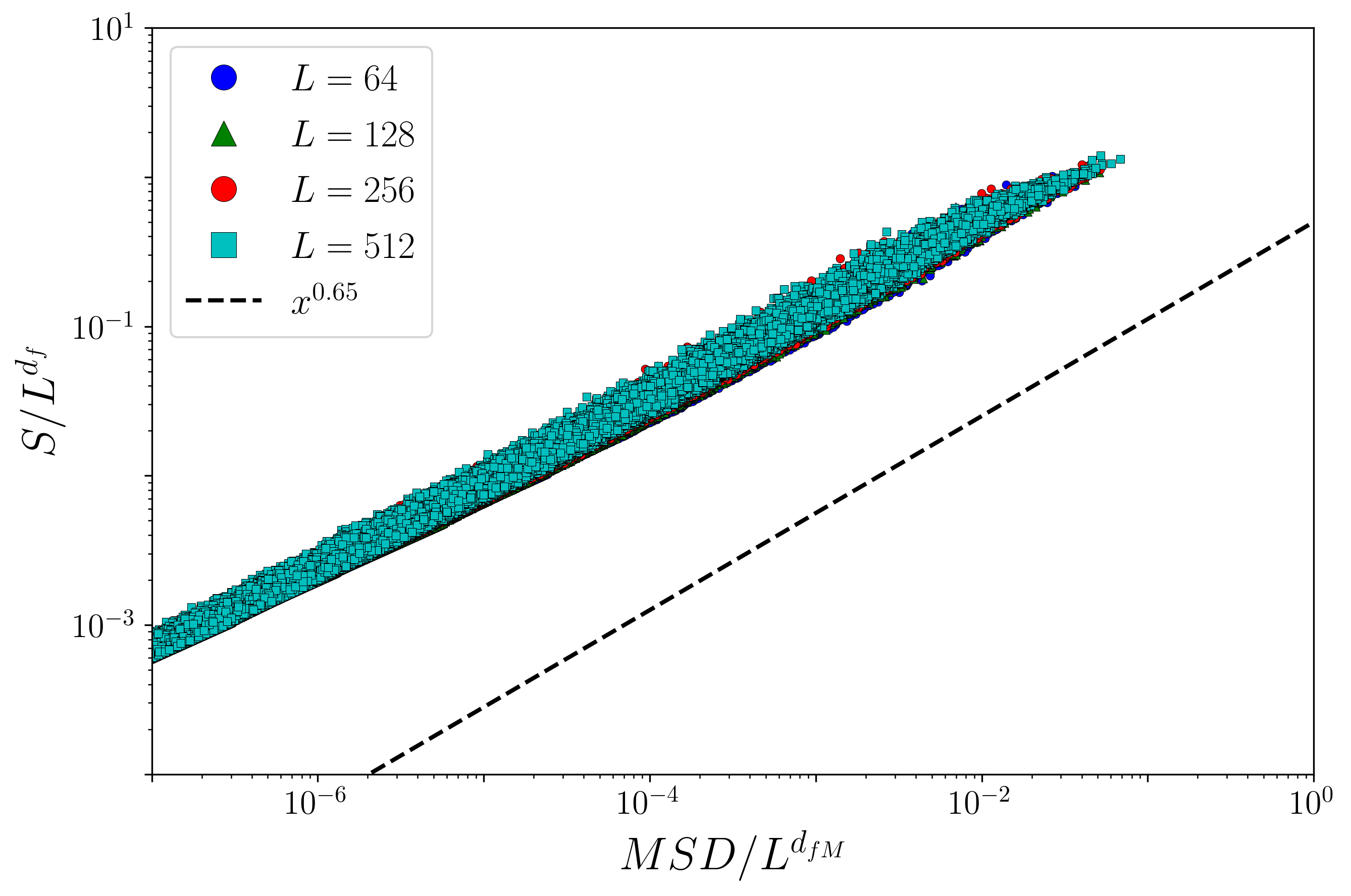} \\
\includegraphics[width=8cm]{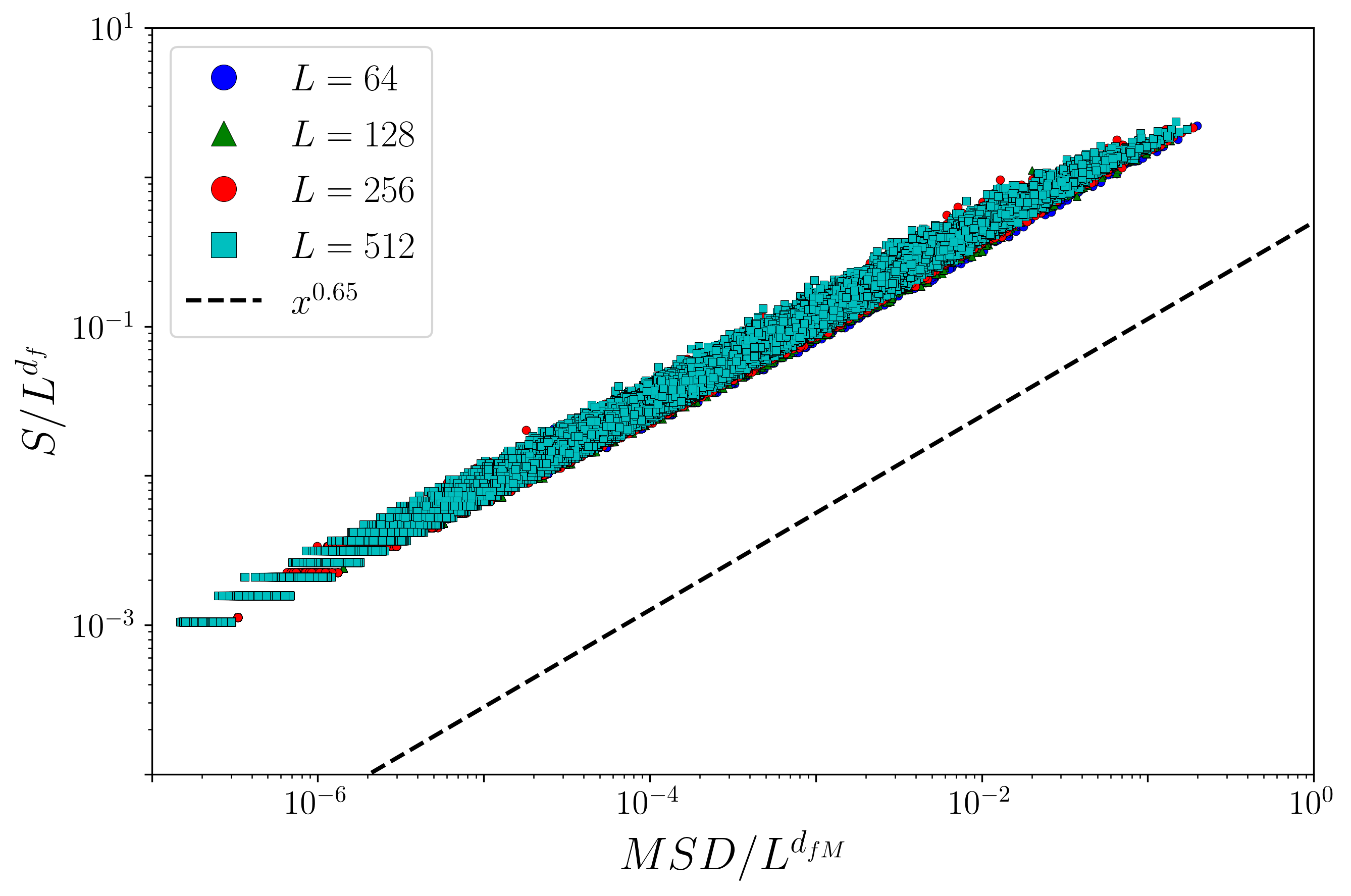}
\caption{\label{fig:individual_avalanches} Representation of individual avalanches. We find an almost one-to-one match between the avalanche size and the corresponding mean square displacement: $S / L^{d_f} \sim (M/L^{d_{fM}})^q$ with $q \approx 0.65$. Top: Y0, bottom: Y1. }
\end{figure}  

As the stress drops and the associated mean square displacements behave in such a similar manner, the question whether we can relate the individual stress drops to the individual mean square displacements naturally arises. Strikingly, we find an almost one-to-one correspondence between the individual avalanche size $S$ and the associated mean square displacement $M$:  $S / L^{d_f} \sim (M/L^{d_{fM}})^q$ with $q \approx 0.65$. Figure \ref{fig:individual_avalanches} shows the $M=M(S)$ dependence: each point represent a single avalanche and they are all narrowly distributed along the $S \sim M^q$ line.  

Starting from this relation and the conservation of the number of events $R(S, L) dS = R_M(M, L) dM$ we can derive two scaling relations:
\begin{eqnarray}
 q(\tau-1) + 1  = \tau_M \\
 d_{fM} [\tau_M - q (\tau+1)] + d_f +\beta = \beta_M
\end{eqnarray}
In the ideal case of perfect slip line avalanches $d_{fM} \to 0$ and the second scaling relation reduces to $d_f + \beta = \beta_M$. We observe, however a slightly larger value, $d_{fM} \approx 0.1$ hence the more complicated scaling form. Similarly, for perfect slip lines one would expect $q=1/2$. Our $q \approx 0.65$ value indicate that avalanches have a more complex structure. Nevertheless, the existence of an $S=S(M)$ relationship completely determines the form of $R(M,L)$ from the distribution $P(S,L)$.

\section{Residual strength statistics}
It has been pointed out previously~\cite{Karmakar2010, Lin2014a} that after a given avalanche, the load increment required to trigger a successive avalanche, $\Delta\gamma$, is completely determined by the weakest site, $\Delta\gamma=\text{min}_i \{x_i\}$, where $x_i=\sigma^y_i-\sigma_i$ where $\sigma^y_i$ is the local value of the yield stress and $\sigma_i$ is the current stress at site $i$.
Under quasistatic loading conditions, at the beginning of an avalanche there is precisely one site with $x=0$ and after an avalanche $x>0$ for all the sites, and the incremental load required to trigger the next event is precisely the minimum value of $x$ at the end of the previous one.
Therefore, one expects a relationship between the distribution of load increments, $P(\Delta\gamma)=P(x_{\text{min}})$, and the distribution of local residual stress values, $P(x)$.

In similar automaton models to ours, Lin et. al. argued that $P(x) \sim x^\theta$ with $\theta \approx 0.6$ as $x \to 0$ \cite{Lin2014a}. 
They used an argument going back to Karmarkar {\it et. al.}~\cite{Karmakar2010} and assumed that $P(\Delta \gamma)$ could be reconstructed from uncorrelated sampling of the $P(x)$ distribution. 
They then used extreme value statistics concepts to relate the size dependence of the mean strain increase between avalanches, $\langle \Delta \gamma \rangle$, to the exponent in the power-law for $P(x)$. 
In extreme value statistics, one has $\int_0^{\langle x_{min} \rangle} P(x) dx \sim 1/N$ where $N$ is the number of uncorrelated samples of $x$ from $P(x)$ and $\langle x_{min} \rangle$ is the average minimum of the $N$-fold sample. 
In ~\cite{Lin2014a} the authors assumed that $P(x)$ had a power law from all the way down to $x=0$ with no finite size effect and from this, they argued that the exponent in the $P(x)$ power law determined the exponent in the size dependence of $\langle \Delta \gamma \rangle = \langle x_{min} \rangle$. 
As we show below, although the relationship we find between $\langle x_{min} \rangle$ and $P(x)$ is consistent with an extreme value statistics argument, the \emph{form} of our $P(x)$ distribution is qualitatively different from Lin {\it et. al.}. 
Rather than seeing a power law down to arbitrarily small $x$, we observe a clear plateau in $P(x)$ at small enough $x$. 
The height of the plateau scales like $1/L^{0.6}$, and the characteristic $x$ value for crossover to the plateau scales like $1/L^{0.9}$ \footnote{As we were preparing the manuscript for submission we discovered that the low $x$ behavior of $P(x)$ had been studied by Jagla and Ferrero~\cite{Ferrero2019} in addition to the PhD thesis of Tyukodi~\cite{Tyukodi2016a}.}
We show below that $\langle \Delta \gamma \rangle = \langle x_{min} \rangle$ occurs at an $x$ value which is not yet on the plateau for any system size we were able to study. 

\begin{figure}[ht]
\begin{center}
\includegraphics[width=6cm]
{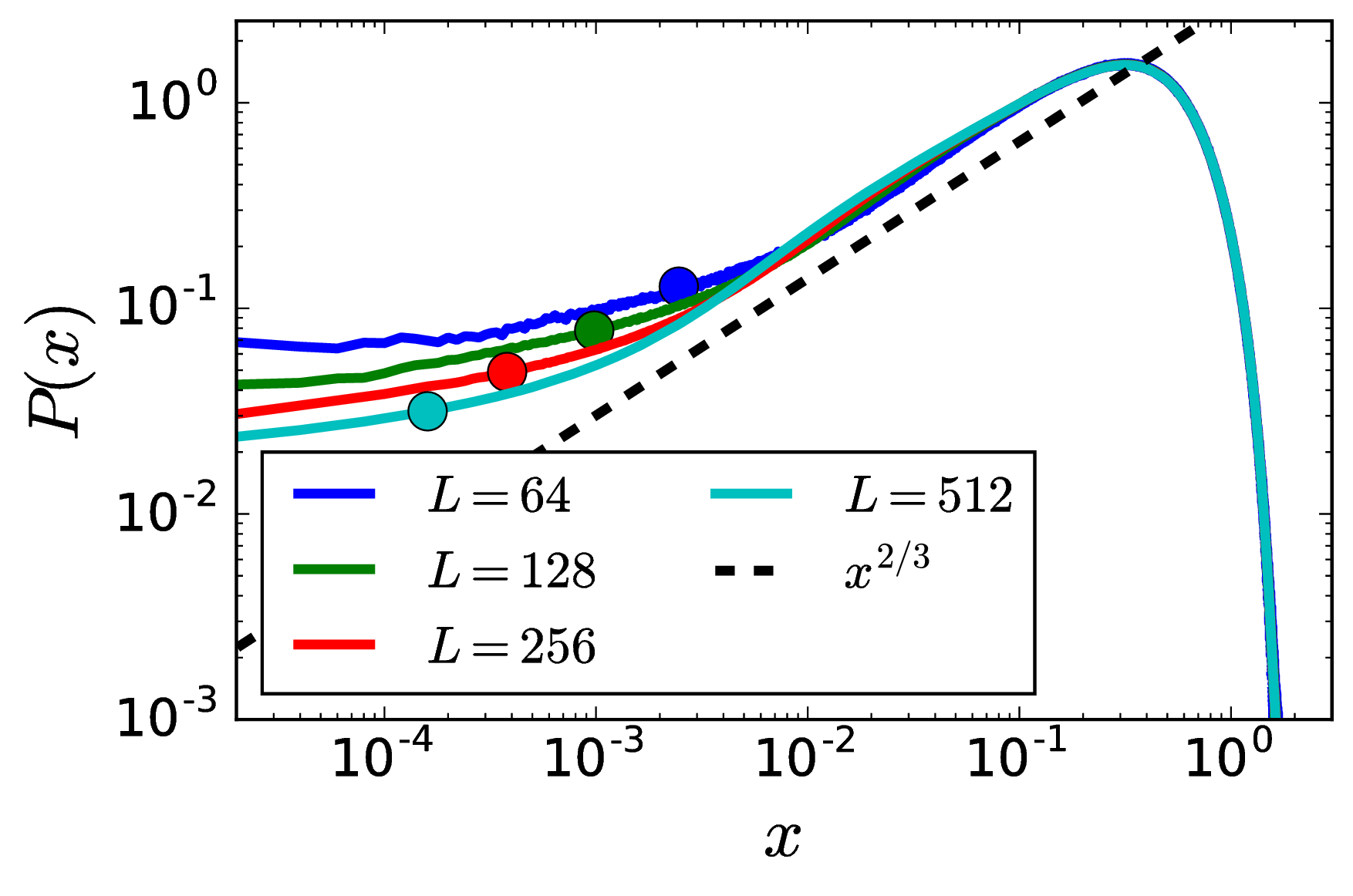}
\includegraphics[width=6cm]
{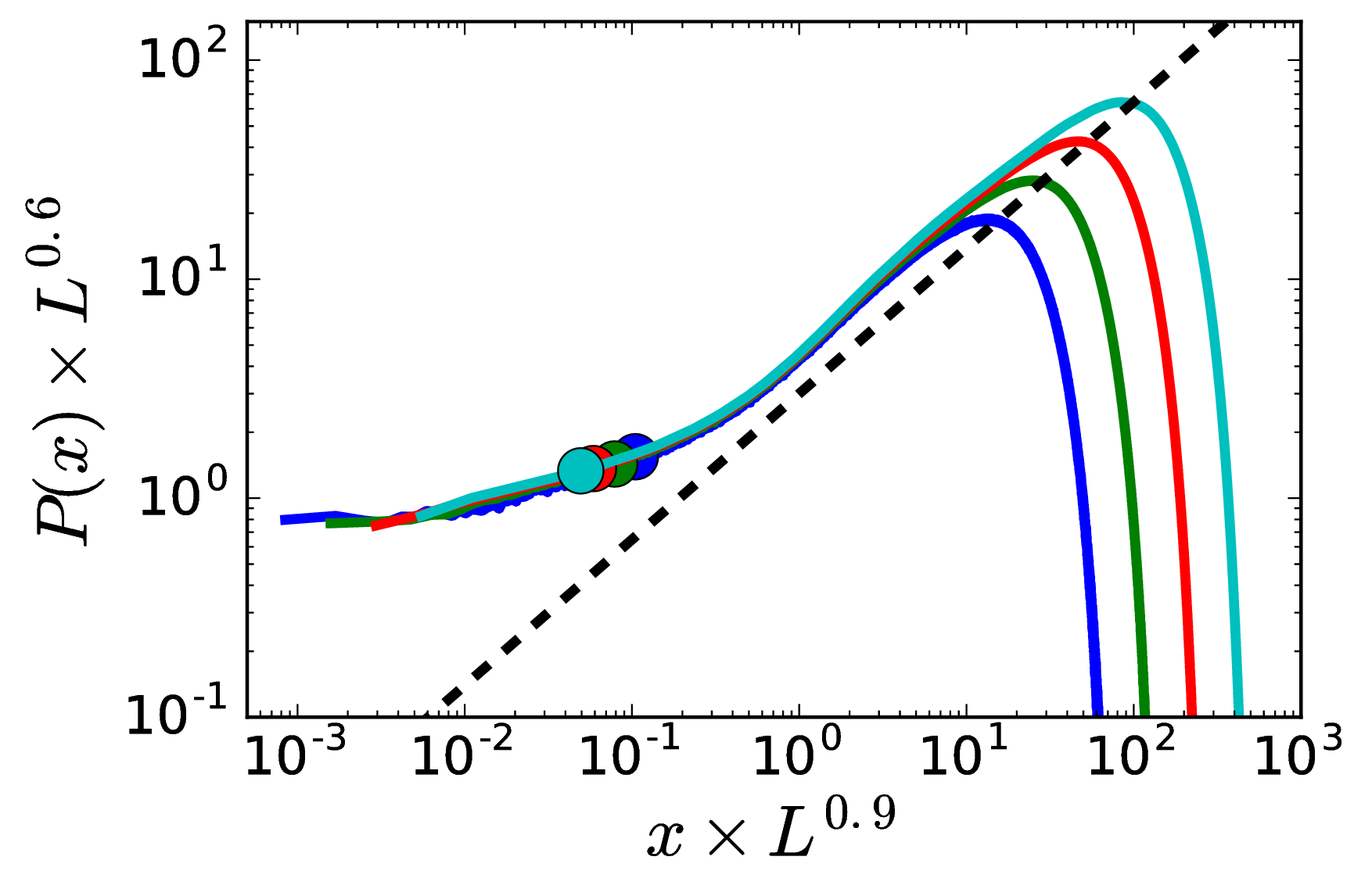}
\caption{\label{fig:Px} $P(x)$ distributions. Note the size dependent lower cutoff. Dots indicate the positions of $\langle  x_{min} \rangle = \langle \Delta \gamma \rangle$ and we observe that they fall off the $P(x) \sim x^\theta$ power law regime.}
\end{center}
\end{figure}

Fig. \ref{fig:Px} shows our measured $P(x)$ distributions. Dots represent positions of $\langle x_{min} \rangle$. It is clear that i) we observe a plateau at the lower end of the distribution and ii) all $\langle x_{min} \rangle$ values lie between the plateau and a power-law like regime. Up until $\langle x_{min} \rangle$ thus $P(x)$ is not a power law. Although the level of the plateau decreases with the system size, so does $\langle x_{min} \rangle$. 

\begin{figure}[ht]
\begin{center}
\includegraphics[width=7cm]
{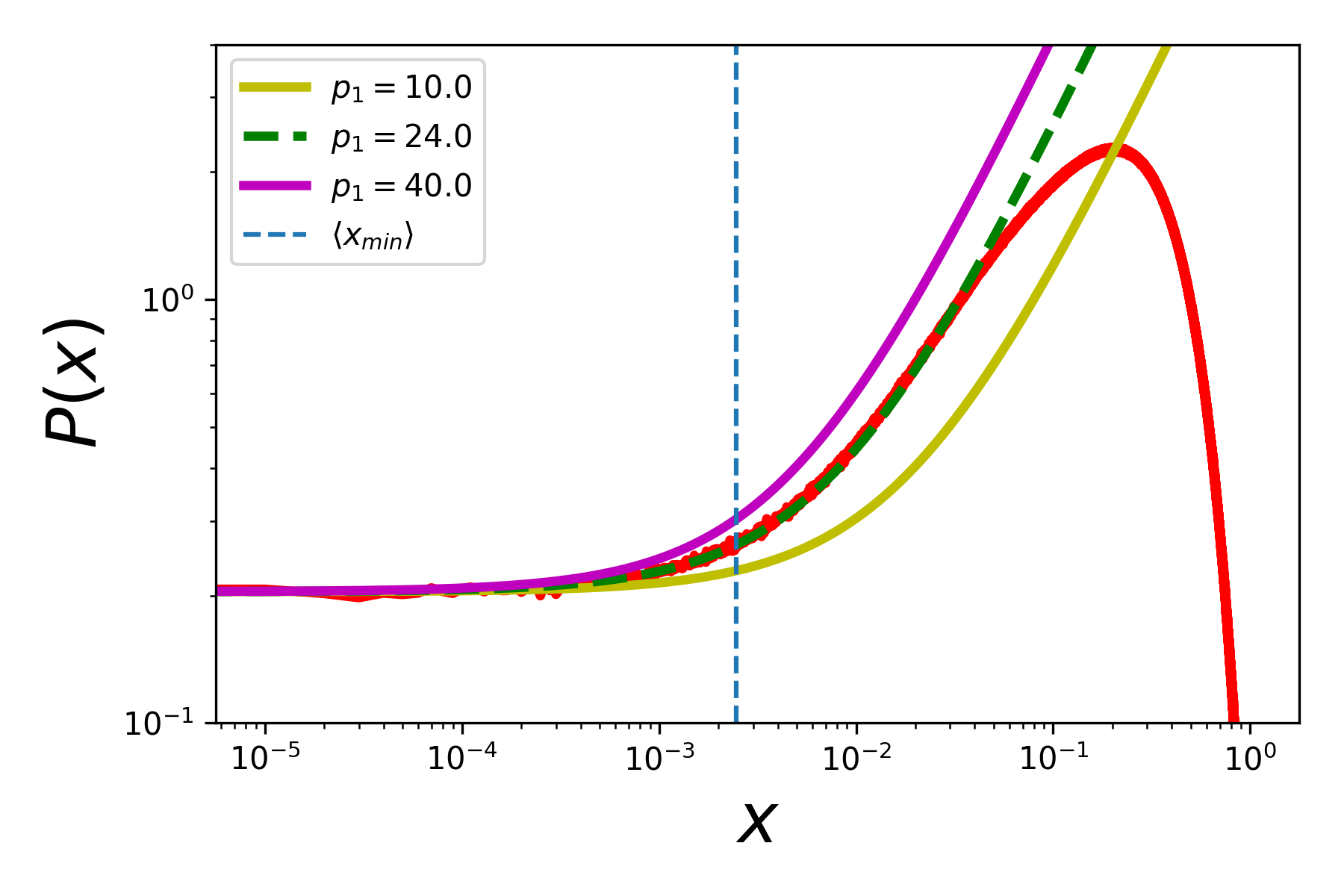}
\caption{\label{fig:Px_lin} Linearized approximation of $P(x)$. Red curve indicates simulation data, the other curves indicate linear approximations $P(x) \approx p_0 + p_1 x$ for various $p_1$ values. }
\end{center}
\end{figure}

Whereas $P(x)$ is not a power law around $\langle x_{min} \rangle$, we still may find a scaling relationship for $\langle x_{min} \rangle$. As $P(x)$ is not a power law, one can approximate it via a simple Taylor expansion $P(x) = p_0 + p_1 x + O(x^2)$ where the coefficients $p_0$ and $p_1$ are both $L$-dependent. We can fix $p_0 = P(x=0)$ and adjust $p_1$ to obtain the best overlap with the measured $P(x)$. Figure \ref{fig:Px_lin} shows the linear approximation of $P(x)$ for various fitting parameters.  

From the collapse on Fig \ref{fig:Px}, it is clear that $p_0 \sim L^{-0.6}$ and $p_1 \sim L^{0.3}$. With these values, we can integrate the linearized $P(x)$ and use the extreme value statistics $\int_0^{\langle x_{min} \rangle} P(x) dx \sim 1/L^2$ to obtain an equation for $\langle x_{min} \rangle$:

\begin{equation}
 p_0 \langle x_{min} \rangle + p_1 \langle x_{min} \rangle^2/2 = c/L^2
\end{equation}

where $c$ is a constant. In the large system limit, one could neglect the $\langle x_{min} \rangle^2$ term, but let us keep it for the moment. Then we have a second order equation in $\langle x_{min} \rangle$. Keeping only the positive solution, we find:
\begin{equation}
 \langle x_{min} \rangle  = (p_0/p_1) [1-(1+2c p_1/(p_0^2 L^2))^{1/2}]
\end{equation}
Using the scaling $p_0 \sim L^{-0.6}$ and $p_1 \sim L^{0.3}$, we get: $\langle x_{min} \rangle  = c' L^{-0.9} [1-(1+2c''L^{-1/2})^{1/2}]$ where $c'$ and $c''$ are two, $L$-independent constants.

In the large system limit, we have 
\begin{equation}
\langle x_{min} \rangle \approx c'L^{-0.9} c'' L^{-0.5} \sim L^{-1.4}
\end{equation}
In the small system size limit, we have
\begin{equation}
\langle x_{min} \rangle \approx c'L^{-0.9}[1 - c'' L^{-0.25}] \sim L^{-1.15}
\end{equation}
The scaling of $\langle x_{min} \rangle \sim L^{-\alpha}$ thus changes with $\alpha \in [1.15, 1.4]$.
For the range of our system sizes, we measure $\alpha \approx 1.35$ (as shown on Fig. \ref{fig:avg_increase}) which is slightly smaller than $1.4$ but note that our $\langle x_{min} \rangle$s are not on the plateau yet and scaling corrections may arise for intermediate system sizes. 

\begin{figure}[ht]
\begin{center}
\includegraphics[width=7cm]
{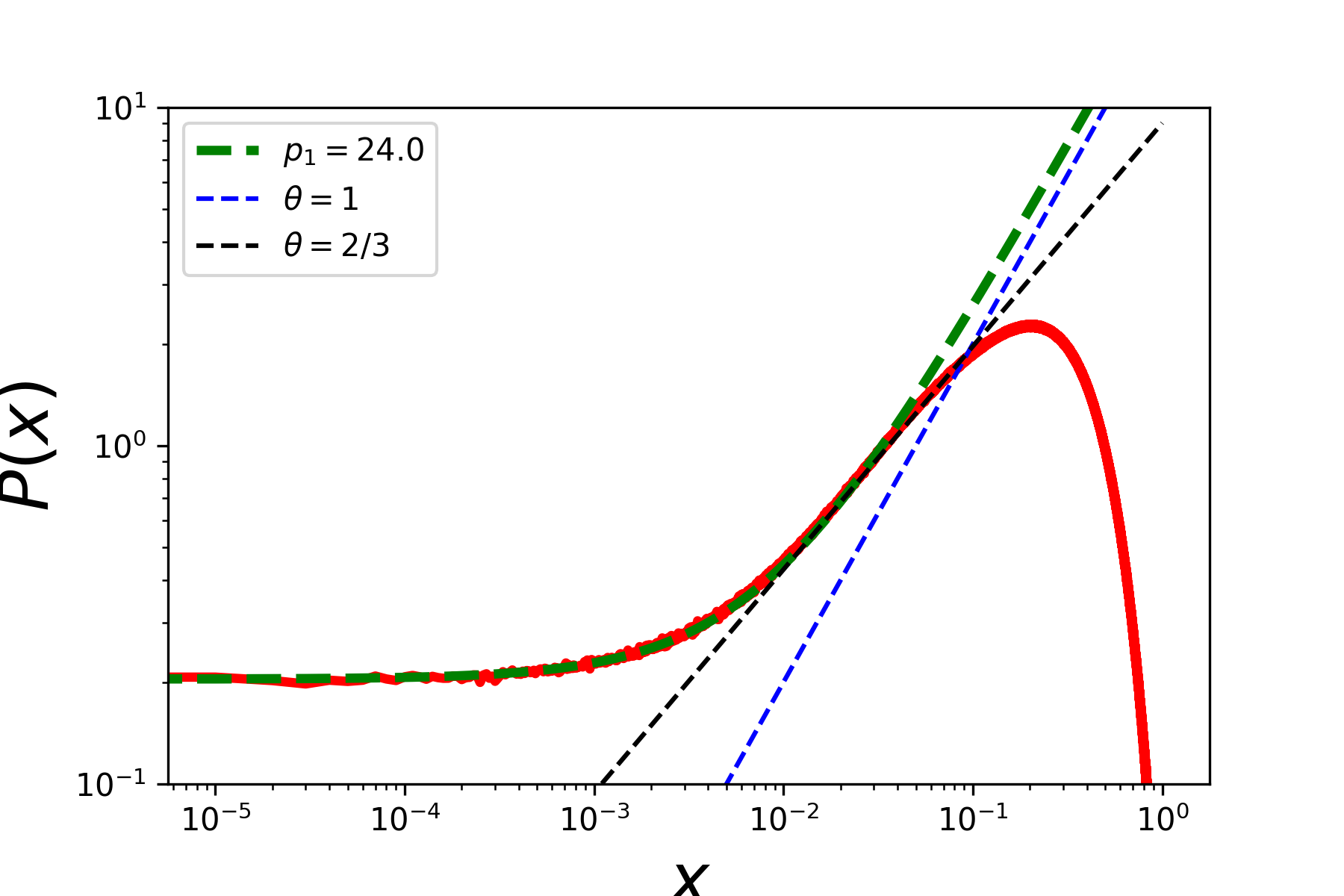}
\caption{\label{fig:Px_lin_PL} Linearized approximation of $P(x)$ appears as a power law with $\theta=2/3$ at intermediate regimes.}
\end{center}
\end{figure}

At large system sizes thus $\langle x_{min} \rangle$ does scale with the system size, even though $P(x)$ is not a power law. There seems to be, however, an intermediate regime where $P(x)$ \emph{appears} as a power law. Figure \ref{fig:Px_lin_PL} shows that even the linearized $P(x)$ appears as a power law $x^\theta$ with $\theta=2/3$. This regime, however, is far above $\langle x_{min} \rangle$, therefore it does not affect the extreme value statistics.

We believe therefore that in truly quasistatic conditions there is an $x_*$ below which $P(x) \to const$ thus there is no so-called pseudogap at $x=0$. Note however, that in finite strain step simulations a natural cutoff comes from the strain step. It is therefore possible that earlier studies with finite strain step or finite rate were only able to observe the strain step imposed cutoff. Furthermore, the initiation of avalanches is different in rigurously quasistatic and finite strain step simulations: in the former, always a single site yields at the triggering moment of an avalanche, while in the latter case there may be multiple sites. Either way, it is not the precise form of $P(x)$ at $x \to 0$ that matters, but rather its form below $\langle x_{min} \rangle$ in determining the scaling of $\langle x_{min} \rangle = \langle \Delta \gamma \rangle$. 

The positions of $\langle x_{min} \rangle$ asymptote toward the plateau as $L$ increases (Fig. \ref{fig:Px}), however, this approach is very slow, and even the $L=512$ system is far above the plateau. This indicates that we have not reached the large system size limit and that one should study extremely large systems in order to avoid corrections to scaling in all quantities. This observation is consistent with scaling corrections to stress drops shown on Fig. \ref{fig:s_mean_scale}.

\section{Diffusion}\label{section:Diffusion}
In the previous section we have been mostly focusing on temporal fluctuations of the stress average, or, equivalently, fluctuations of the elastic strain and plastic strain \emph{averages}, but showed that the spatial structure of these avalanches has a fractal dimension $d_f \approx 1.1$, suggesting that avalanches have an almost linear shape. 
The consequence of such anisotropic avalanches is a highly nonhomogeneous strain/displacement field that is impossible to capture by usual depinning or mean field models. More importantly, instead of saturating to a steady state value, spatial inhomogeneities (i.e. fluctuations) in the strain/displacement fields keep increasing with time. A diffusive increase of the displacement fluctuations (i.e. mean square displacement) was observed in several particle simulations \cite{Maloney2008, Lemaitre-PRE07, Lemaitre2009, Roy2015, Tsamados-PhD09}, as well as in lattice models \cite{Martens2011a}. Moreover, a monotonic increase of the variance of the incremental plastic strain field with the window size, $\Delta \epsilon$, was previously reported \cite{TPRV-PRE16, TPVR-Meso12} in lattice models similar to the one we study here. We note that, although we find a regime in window size, $\Delta \epsilon$, where the effective diffusion coefficient is essentially constant, the displacement fields which give rise to this diffusion coefficient are strongly spatially correlated.  Furthermore, the displacement distribution is strongly non-Gaussian in this regime. This is in stark contrast with diffusion in simple liquids where there is relatively little displacement correlation between initially neighboring tracer particles on the diffusive time scale and the displacement distribution becomes completely Gaussian.  Here, rather, the linear evolution of the second moment of the displacement distribution arises from single avalanches occurring within a given window of size $\Delta \epsilon$.  As the window size grows to encompass multiple avalanches, inter-avalanche correlations come into play, and the second moment of the displacement distribution becomes super-diffusive.



In ref. \cite{Tyukodi2018} we showed that diffusion occurs in mesomodels and the finite size scaling of the diffusion coefficient is consistent with particle simulations. We found that there is a short-term and a long-term diffusive as long as the kernel is constructed properly. Here we extend the analysis and show that the short-time diffusive behavior is a result of the shotnoise of uncorrelated avalanches and we connect the finite size scaling of the diffusion coefficient $D$ to the finite size scaling of the per-avalanche mean square displacement and of the load increment.

\subsection{Diffusive increase of plastic strain field fluctuations}

\begin{figure}[ht]
\begin{center}
\includegraphics[width=8cm]
{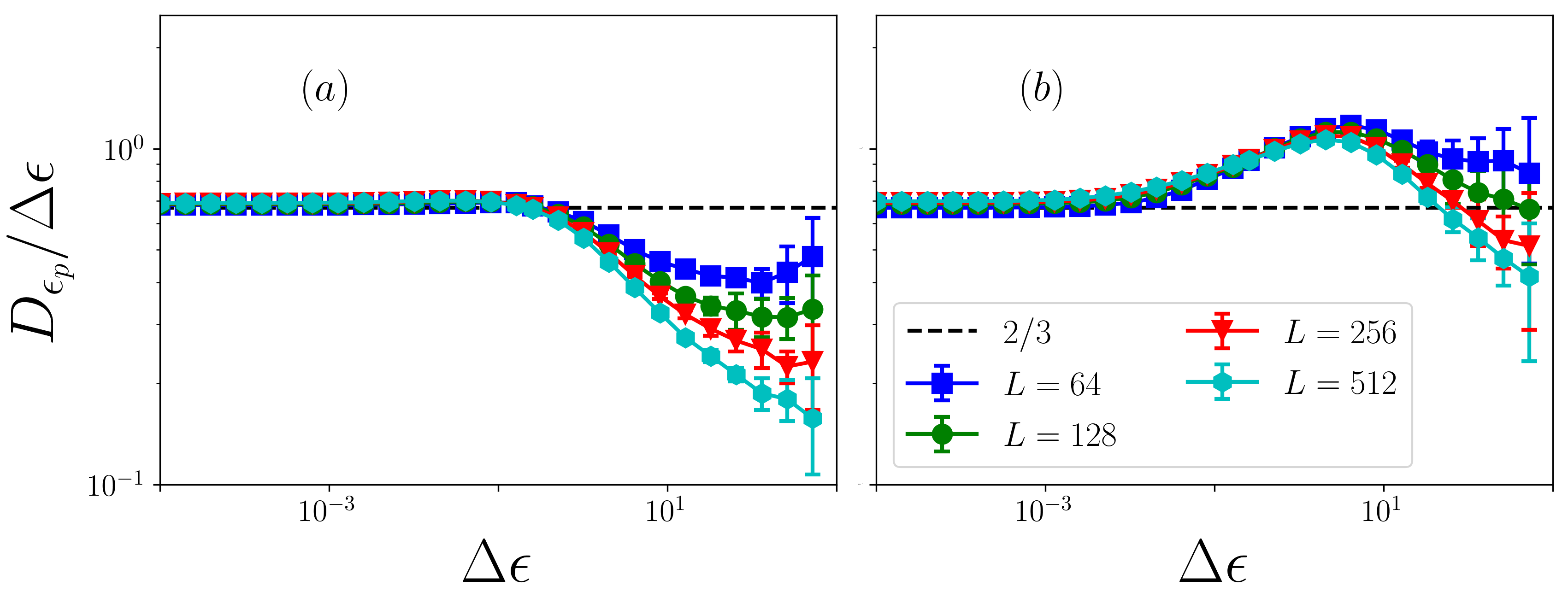}
\caption{\label{fig:dep2} Plastic strain diffusivities for (a) Y0 and (b) Y1, various system sizes. We observe a short time, size independent diffusivity and a slow convergence to a second diffusive regime.}
\end{center}
\end{figure}

Figure \ref{fig:dep2} shows the plastic strain diffusivity $D_{\epsilon_p} = \langle \delta \epsilon_p^2 \rangle / \Delta \epsilon$ for various system sizes. Here $\langle \delta \epsilon_p^2 \rangle$ is the plastic strain field's variance where the plastic strain was accumulated over a window of size $\Delta \epsilon$. For both protocols, we observe a clear diffusive behavior $D_{\epsilon_p} = const$ for short windows, i. e. $\Delta \epsilon < \epsilon_0/2$. Moreover, the associated diffusivity $D_{\epsilon_p}$ is size independent for short times and we find $D_{\epsilon_p} \approx 2/3$. This value can be understood assuming that the probability distribution of plastic strains is a uniform distribution corresponding to sites which have yielded precisely once and zero to the sites which have not yielded. This simple estimate gives a value $2/3$.

After a strain of $\Delta \epsilon \approx \epsilon_0/2$ however, we observe a departure from diffusive behavior and a slow convergence to a second diffusive regime.

The fall-off from the initial $D_{\epsilon_p}=const$ plateau in both cases happens at around $\Delta \epsilon \approx \epsilon_0/2$ which is precisely the strain necessary for each site to yield once. Since the plastic strain field is local, this is an indication that temporal correlations only start to build up after each site in the system yielded on average and then the long-term behavior is a result of temporal correlations in the plastic activity.

\subsection{Diffusive increase of mean square displacements}
In particle simulations there is an ambiguity in separating local strain fields into elastic and plastic parts and one has more direct access to the non-affine displacements of particles. For a more straightforward comparison to particle simulation results therefore we investigate the fluctuations of the displacement fields.

Figure \ref{fig:du2_1} shows the diffusion coefficients $D = \langle \delta u^2 \rangle / \Delta \epsilon$ for the two protocols, where $\langle \delta u^2 \rangle$ is the mean square displacement of the displacement field accumulated over a window of size $\Delta \epsilon$. At short times, we again observe diffusion, followed by a crossover to a superdiffusive regime and then a second diffusive behavior.

\subsubsection{Early diffusion}

\begin{figure}[ht]
\begin{center}
\includegraphics[width=9cm]
{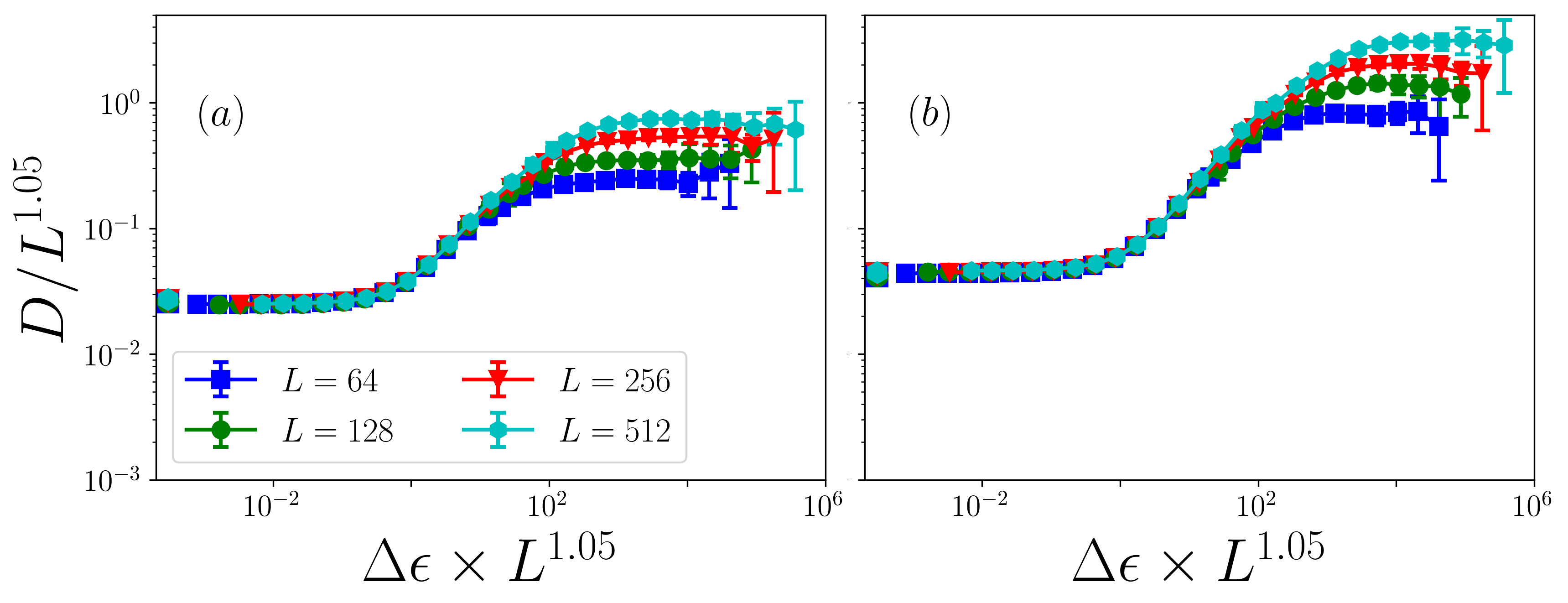}
\caption{\label{fig:du2_1} Diffusion coefficients for various system sizes, (a) Y0 and (b) Y1. At short times, we find $D \sim L^{1.05}$ for both protocols. The leftmost dots indicate predictions of the diffusion coefficient $D = \langle MSD \rangle / \langle \Delta \gamma \rangle$}
\end{center}
\end{figure}

At short times, in contrast with the plastic strain diffusivity, $D$ is system-size dependent and we find $D \sim L^{1.05}$ so $D/L^{1.05}$ gives a good collapse for various system sizes (Fig. \ref{fig:du2_1}). After a characteristic strain $\epsilon^* \sim L^{1.05}$ however, we observe a departure from this initial diffusive plateau.

\onecolumngrid

\begin{figure}[h] 
\includegraphics[width=16cm]{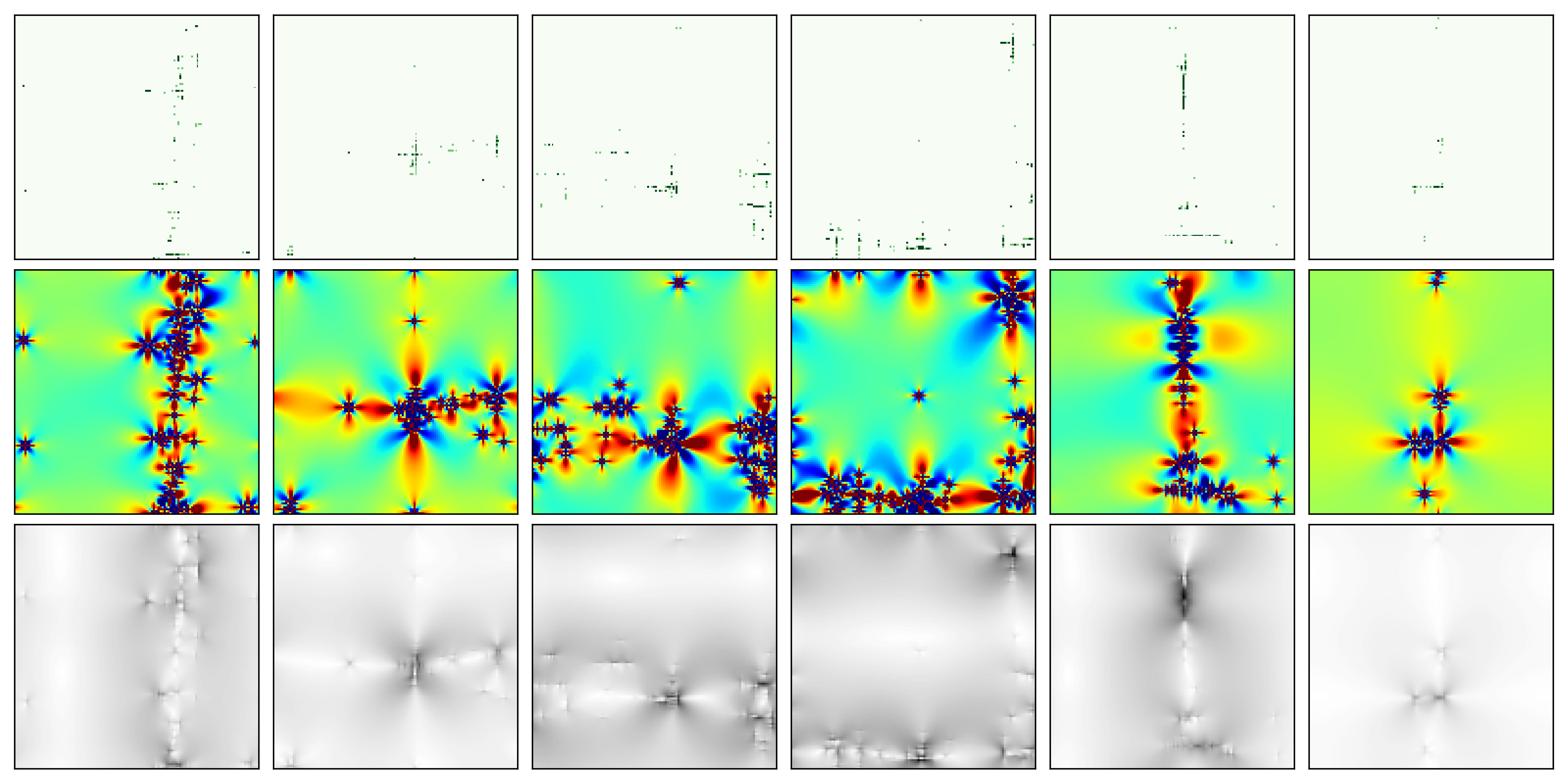}
\caption{\label{fig:sliplines} Slipline formation. Maps from top to bottom: plastic strain, stress, displacement modulus $|u|$, for successive strain windows of size $\epsilon_0/L$. Such a strain window, on average, allows for the formation of a single slip.}
\end{figure}  

\begin{figure}[h] 
\includegraphics[width=16cm]{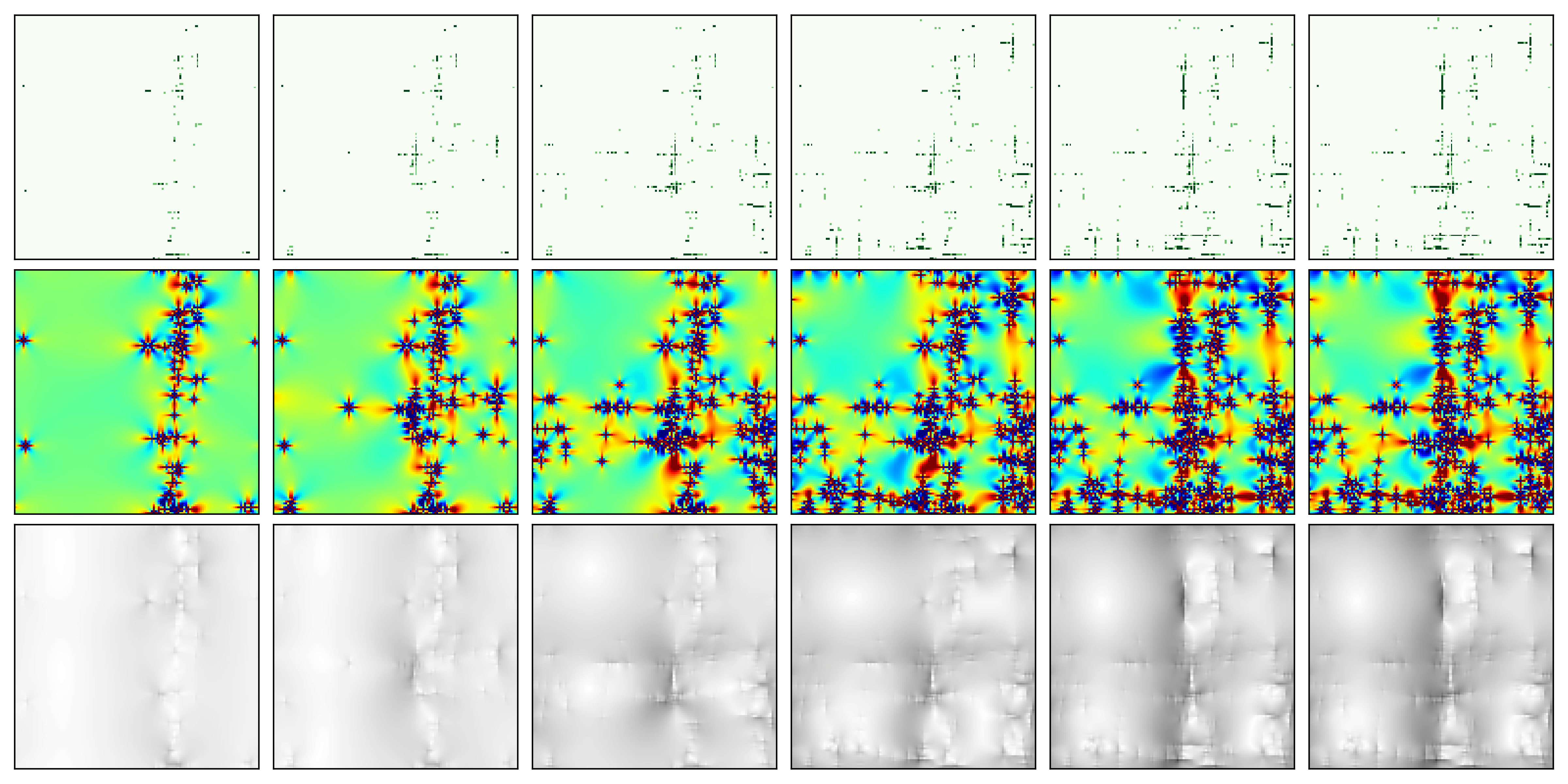}
\caption{\label{fig:sliplines_cumm} Slipline formation. Cummulative maps, accumulated in successive increments of strain window of size $\epsilon_0/L$. Maps from top to bottom: plastic strain, stress, displacement modulus $|u|$. Maps correspond to snapshots from Fig. \ref{fig:sliplines}}
\end{figure}  

\twocolumngrid

Our size-scaling of $D$ is very close to $D \sim L$ found in several independent particle simulations \cite{Maloney2008, Lemaitre-PRE07, Lemaitre2009, Roy2015, Tsamados-PhD09} and it has been associated to the formation of system-spanning slip-lines. According to the ``perfect slip-line'' hypothesis, avalanches have a spatial structure of perfect, system spanning slip lines. There are $L/a$ flipping sites on the line, each of them having an average plastic strain $\epsilon_0/2$. The plastic strain accumulated by such a slip line is then given by $\epsilon_s = a \epsilon_0/2L$, whereas the displacement field variance associated to a line is \cite{Tyukodi2018} $\langle \mathbf{u_s}^2 \rangle = a^2 \epsilon_0^2 / 12$. In the steady state, the stress cannot increase nor decrease on average, therefore elastic strains must be equal to the plastic strains. The diffusion coefficient is then given by $D = \langle \mathbf{u_s}^2 \rangle / \epsilon_s = a \epsilon_0 L / 6 \sim L$. 

Figure \ref{fig:sliplines} shows the plastic strain, stress and displacement fields accumulated over subsequent windows of size $\Delta \epsilon = \epsilon_0 / 2 L$ which is precisely the strain necessary for one slip line to form. We observe that within these windows most of the time there is one slip line forming, these lines however are not perfect. 

We can extend the above argument by, instead of considering perfectly linear objects, assigning avalanches a non-trivial fractal dimension $d_f$. In this case, the plastic strain released by one avalanche is $\epsilon_s = (L^{d_f} / L^2) a \epsilon_0/2 = (a \epsilon_0/2) L^{d_f-2}$. Assuming that $\langle \mathbf{u_s}^2 \rangle$ is still size independent, one finds $D \sim L^{2-d_f}$. Comparing against our observation $D \sim L^{1.05}$ we can infer a fractal dimension $d_f = 0.95$. While this value is consistent with particle simulations \cite{Salerno2012a} and other lattice models \cite{Talamali2011a}, it is inconsistent with our direct measurement of the avalanche cutoff finite size scaling where we found $d_f \approx 1.1$ indicating that the mean square displacement resulting from one of the fractal objects has a slight system-size dependence. Indeed, as we show on Figure \ref{fig:P_MSD} inset, the average $MSD$ of avalanches has a weak size dependence: $\langle MSD \rangle \sim L^{-0.28}$. The reason of this size dependence can be two-fold: first, events in avalanches may not be perfectly aligned along a line, second, multi-flips may occur which gives an extra, plastic strain/flip number dimension to the avalanche shapes. Note that the existence of diffusion does not require any particular spacial structure of the individual avalanches. What happens is simply that in most windows no avalanche occurs and in a small number of windows a single avalanche occurs. The mean square displacement at a particular window size thus is a weigthed average of a large number of zero values resulting from the windows with no events and a small number of nonzero values resulting from individual avalanches. The initial diffusive behavior is thus a result of the shotnoise of individual avalanches.

The diffusion coefficient is then given by 
\begin{equation}\label{eq:D}
D = \frac{\langle MSD \rangle}{\langle \Delta \gamma \rangle}
\end{equation}

This relation is verified and shown on Fig \ref{fig:du2_1}: the initial dots indicate the prediction of $D$ from the $MSD$ and $\Delta \gamma$ measurements. Equivalently, $\langle MSD \rangle = D \langle \Delta \gamma \rangle$. Using $D \sim L^{1.05}$ and $\langle \Delta \gamma \rangle \sim L^{-1.35}$ we find that $\langle MSD \rangle \sim L^{-0.3}$, a scaling relation connecting the size dependence of the elementary events to the size dependence of the diffusion coefficient. The relation is supported by our data, as shown on Figure \ref{fig:P_MSD} inset.

\subsubsection{Crossover to superdiffusive scaling and long term diffusion}
\begin{figure}[ht]
\begin{center}
\includegraphics[width=8cm]
{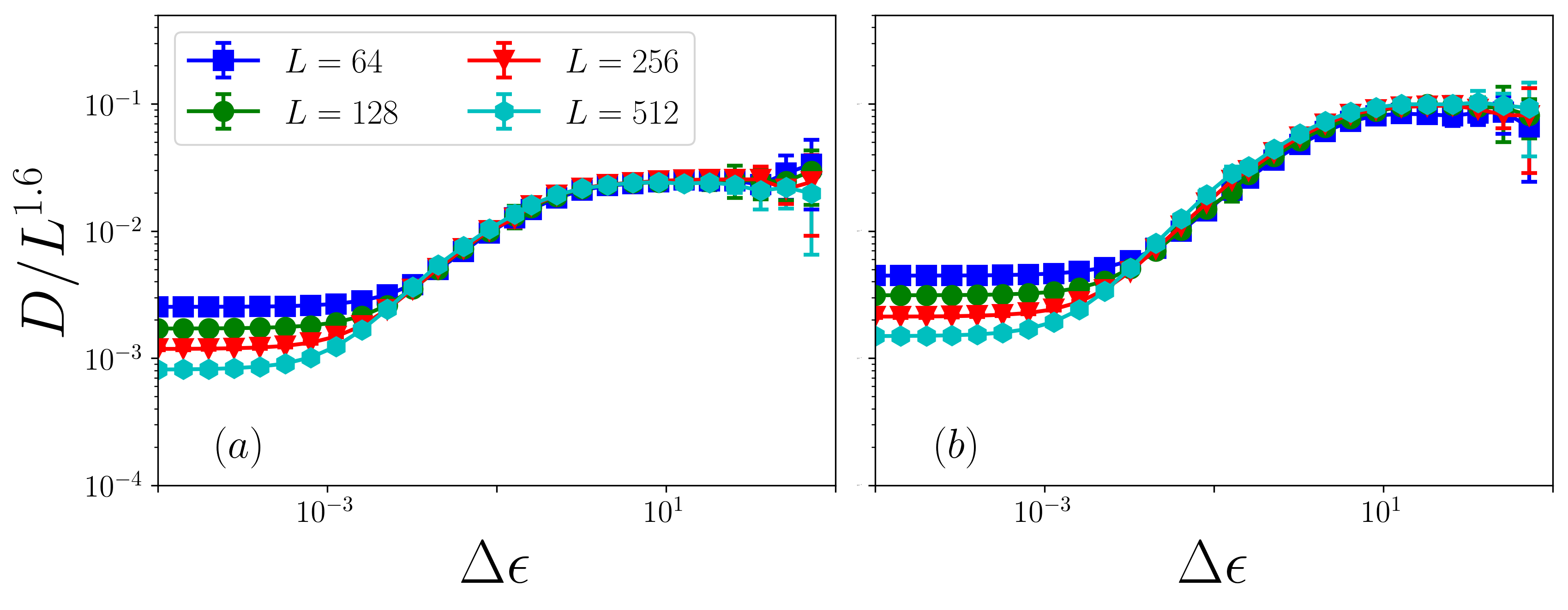}
\caption{\label{fig:du2_1p6} Diffusion coefficients for (a) Y0 and (b) Y1. We find a long time diffusive regime with $D \sim L^{1.6}$ independently of the protocol.}
\end{center}
\end{figure}

The simple minded picture of individual slip-lines only holds up to a strain $\epsilon^* \sim L^{-1.05}$ which is precisely the strain necessary for a single slip line to form. Passed this strain we observe a supperdiffusive increase of the mean square displacement as shown in Fig. \ref{fig:du2_1p6}, indicating that a correlation starts to build up between subsequent slip lines. The build-up of this correlation can be observed in the strain, stress and displacement fields as well: in Fig. \ref{fig:sliplines_cumm} we show the same snapshots as in Fig. \ref{fig:sliplines}, this time however accumulating deformation from the first window. As slip lines add up, it is clear that they are not independent. Note however, that we never observe persistent localization: although the decorrelation time (i. e. the time required for the plastic activity to leave a band and move to another one) increases with time, plastic activity will eventually decorrelate.

At long enough times, the system reaches a second diffusive regime. The diffusion coefficient now has a different scaling, we find $D \sim L^{1.6}$ for this late diffusive regime, which is consistent with Martens et al. \cite{Martens2011a} where $D \sim L^{1.5}$ was reported in a similar lattice model.

\begin{figure}[ht]
\begin{center}
\includegraphics[width=7cm]
{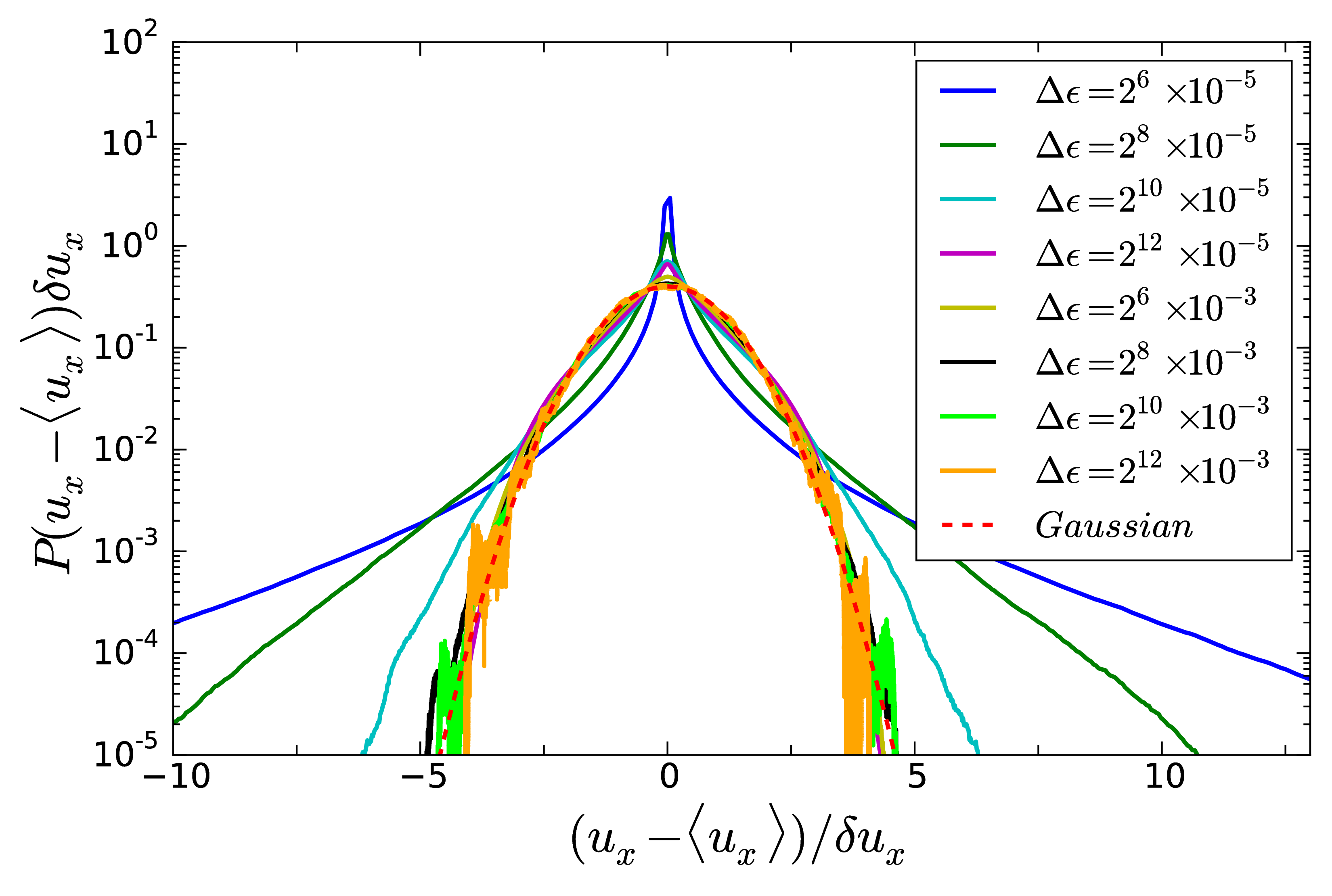}
\includegraphics[width=7cm]
{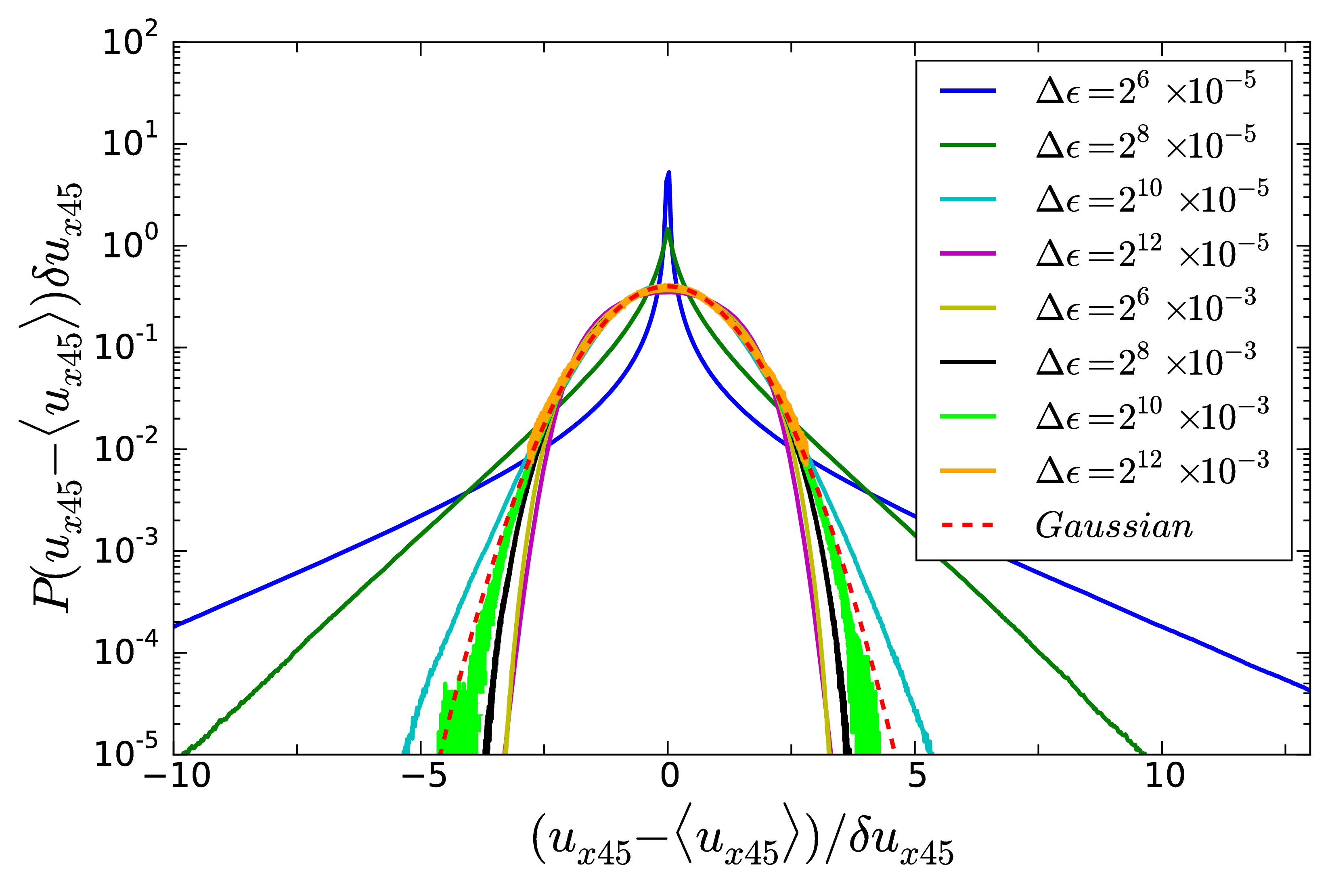}
\caption{\label{fig:hist_ux} Distributions of the cartesian components of the displacements, $P(u_x)$ and $P(u_{x45})$ for Y1, increasing strain windows.}
\end{center}
\end{figure}

\begin{figure}[ht]
\begin{center}
\includegraphics[width=9cm]
{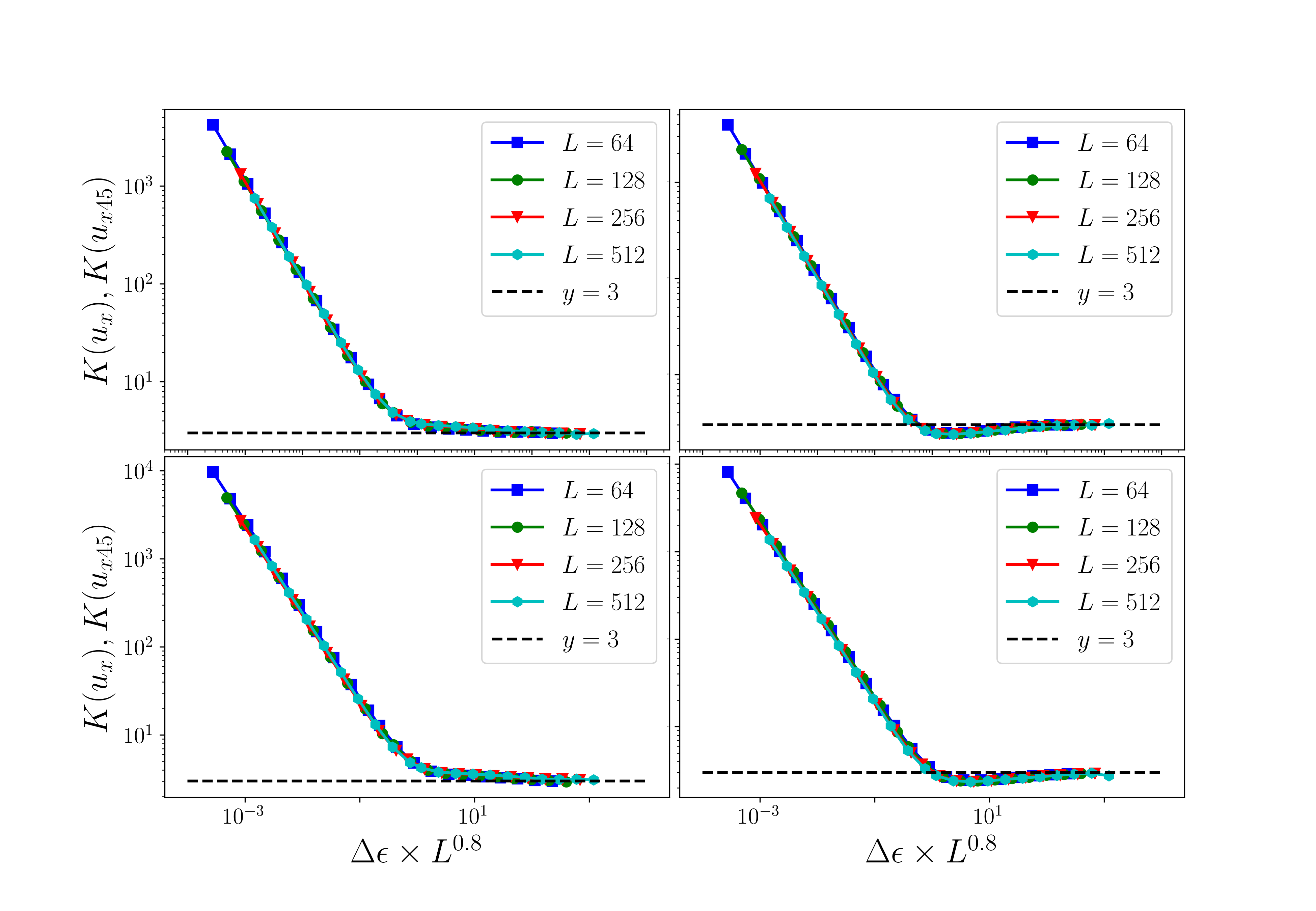}
\caption{\label{fig:kurtosis_y0} Kurtosis for Y0 (top) and Y1 (bottom), axial (left) and diagonal (right). The two protocols are indistinguishable.}
\end{center}
\end{figure}

As we have argued previously, the linear increase of fluctuations with time does not imply a one-particle diffusion process. Figure \ref{fig:hist_ux} shows the evolution of the distributions of the $u_x$ and $u_{x45} = (u_x+u_y)/\sqrt{2}$ components of the displacements. For short times, i.e. $\Delta \epsilon < \epsilon^*$ we observe a distribution with an exponential tail which is the signature of the displacement field induced by individual slip lines \cite{Maloney2008, Tsamados2009a, Tsamados-PhD09}. Considering the distributions only, it may be tempting to model the evolution of displacements as a random walk with exponentially distributed steps. While such an approach indeed predicts diffusion and complies with the observed displacement distributions (exponential tail at short time and normal distribution at long times), it does not account for the size effects caused by the localization described above. Chaudhuri, Berthier and Kob have argued that exponential tails arise generically in the displacement distributions of glassy systems using a continuous time random walk framework~\cite{Chaudhuri2007}.  There, the exponential tails are populated with particles which have undergone a larger number of discrete CTRW jumps than average.  Here, although we observe exponential displacement distributions, the origin is completely different.  For the earliest times, a site has essentially undergone either one (with probability proportional to $\langle \Delta \gamma \rangle/\Delta\epsilon$) or zero avalanches and the occurrence of multiple jumps is exceedingly rare ($\leq \Delta\epsilon^2$).  The exponential displacement distribution is a consequence of the spatial structure of the displacement fields which arise from single avalanches.  This is a completely different scenario than the CTRW proposed in~\cite{Chaudhuri2007}.

At long times, the distributions of the cartesian components of the displacement field converge to a normal distribution and the distribution of its magnitude to a Maxwell distribution. The variance of the normal distribution then increases linearly with time. Figure \ref{fig:kurtosis_y0} quantifies the convergence to a normal distribution by following the evolution of the kurtosis of the displacement distributions. The kurtosis $K$ of the $u_x$ and $u_{x45}$ cartesian components of the displacement field shows an initial $K \sim 1/\Delta \epsilon$ decrease. This behavior can be understood in terms of shotnoise avalanches \cite{Tyukodi2018}: all moments of the distribution should scale as $\langle \delta u^n \rangle \sim \Delta \epsilon$, thus, for the kurtosis we have $\langle \delta u^4 \rangle / \langle \delta u^2 \rangle^2 \sim \Delta \epsilon / \Delta \epsilon^2 = 1/\Delta \epsilon$. At long times, we recover $K\approx 3$ indicating a normal distribution of displacement components.

\section{Conclusions}\label{section:Conclusion}
In summary, we have studied three different aspects of a meso-scale automaton model for a-AQS systems:  i) event and inter-event statistics, ii) residual threshold distribution, iii) diffusion and have shown that all three are inter-related.
The average $\langle S\rangle$ of the distribution, $P(S)$, of stress drops completely determines the distribution of load increments $P(\Delta \gamma)$ under the assumption that the latter is a simple exponential.  
The distribution, $P(M)$, of single-event MSDs along with the average load increment, $\langle \Delta \gamma \rangle$, completely determine the effective diffusion coefficient $D=\langle M \rangle / \langle \Delta\gamma \rangle$. 
The $P(M)$ distribution is determined completely from the $P(S)$ distribution and a single scaling relation between $S$ and $M$, $M\propto S^{q}$ with $q\approx 0.65$.

The distribution of residual strengths, $P(x)$ was found to be analytic at $x\rightarrow 0$.
The value at $x=0$ was found to scale like a power of the system size in a way which is consistent with what would be predicted from $\langle \Delta \gamma \rangle$ along with extreme value statistics arguments.
While our particular finding on the form of $P(x)$ is different from that found by Lin {\it et. al.}\cite{Lin2014a}, we nonetheless find it likely that the basic extreme value argument first put forward by Karmarkar {\it et. al.}~\cite{Karmakar2010} is essentially correct.
We showed that our analytic for for $P(x)$ shows an apparent power-law regime with an exponent consistent with that measured by Lin, however, this apparent power-law regime occurs at $x$ values which are well above $x_{min}$ and should have no impact on $\langle \Delta \gamma \rangle$.

For the diffusion coefficient, as we have shown previously~\cite{Tyukodi2018}, there is an early time diffusive regime and a late time diffusive regime with a higher diffusion coefficient.
The diffusion coefficient for the early time regime is the one which is precisely $\langle M \rangle / \langle \Delta \gamma \rangle$ and therefore intimately related to the avalanches and residual thresholds.
The diffusion coefficient increases beyond the early-time plateau value at a characteristic strain which scales with system size in precisely the same way as the height of the plateau itself.
We have discussed the connection between the height of the early time plateau and the avalanches, but it is not completely clear to us why the characteristic strain for departure from the plateau scales in precisely the same way with system size as the height of the plateau itself.
We are content here to leave it as an empirical observation, but it deserves further study in the future.
In a subsequent paper, we will also more fully study the late time diffusive regime.

\begin{acknowledgments}
This material is based upon work supported by the National Science Foundation under Grants CMMI-1822020 and PHY-1748958. \textcolor{white}{\cite{Ferrero2019}}

\end{acknowledgments}

\bibliographystyle{aipauth4-1}
\bibliography{library}
\end{document}